\documentclass[useAMS, onecolumn]{mn2e}
\pdfoutput=1
\usepackage{epsfig}
\usepackage{amsmath}
\usepackage{amssymb}
\def\be{\begin{equation}}
\def\ee{\end{equation}}
\def\ba{\begin{eqnarray}}
\def\ea{\end{eqnarray}}
\def\go{\mathrel{\raise.3ex\hbox{$>$}\mkern-14mu
             \lower0.6ex\hbox{$\sim$}}}
\def\lo{\mathrel{\raise.3ex\hbox{$<$}\mkern-14mu
             \lower0.6ex\hbox{$\sim$}}}
\def\bxi{{\mbox{\boldmath $\xi$}}}
\def\br{{\bf r}}

\def\sun{\odot}
\newcommand{\numax}{\mbox{$\nu_{\rm max}$}}

\newcommand{\muHz}{\mbox{$\mu$Hz}}

\begin{document}

\title[Tides in Triple Systems]{Tidally Induced Oscillations and Orbital Decay in Compact Triple-Star Systems}

\author[J. Fuller et al.] {Jim Fuller$^1$\thanks{Email: derg@astro.cornell.edu},
A. Derekas$^{2,3}$, T. Borkovits$^{4,2,6}$, D. Huber$^{3,5}$, T. R. Bedding$^3$, L. L. Kiss$^{2,3,6}$
\\$^1$Center for Space Research, Department of Astronomy, Cornell University, Ithaca, NY 14853, USA
\\$^2$Konkoly Observatory, Research Centre for Astronomy and Earth Sciences, Hungarian Academy of Sciences,
 H-1121 Budapest, \\ Konkoly Thege Mikl\'os \'ut 15-17, Hungary
\\$^3$Sydney Institute for Astronomy, School of Physics, University of Sydney, NSW 2006, Australia
\\$^4$Baja Astronomical Observatory, H-6500 Baja, Szegedi \'ut, Kt. 766, Hungary
\\$^5$NASA Ames Research Center, Moffett Field, CA 94035, USA
\\$^6$ELTE Gothard-Lend\"ulet Research Group, H-9700 Szombathely, Szent Imre herceg \'ut 112, Hungary
}

\label{firstpage}
\maketitle
%%%%%%%%%%%%%%%%%%%%%%%%%%%%%%%%%%%%%%%%%%%%%%%%%%%%%%%%%%%%%%%%%%%%
\begin{abstract}

We investigate the nature of tidal effects in compact triple-star systems. The hierarchical structure of a triple system produces tidal forcing at high frequencies unobtainable in binary systems, allowing for the tidal excitation of high frequency p-modes in the stellar components. The tidal forcing exists even for circular, aligned, and synchronized systems. We calculate the magnitude and frequencies of three-body tidal forcing on the central primary star for circular and coplanar orbits, and we estimate the amplitude of the tidally excited oscillation modes. We also calculate the secular orbital changes induced by the tidally excited modes, and show that they can cause significant orbital decay. During certain phases of stellar evolution, the tidal dissipation may be greatly enhanced by resonance locking. We then compare our theory to observations of HD~181068, which is a hierarchical triply eclipsing star system in the Kepler field of view. The observed oscillation frequencies in HD~181068 can be naturally explained by three-body tidal effects. We then compare the observed oscillation amplitudes and phases in HD~181068 to our predictions, finding mostly good agreement. Finally, we discuss the past and future evolution of compact triple systems like HD~181068.

\end{abstract}

\begin{keywords}
stars: oscillations --- stars: multiple --- stars: individual: HD 181068
\end{keywords}

%%%%%%%%%%%%%%%%%%%%%%%%%%%%%%%%%%%%%%%%%%%%%%%%%%%%%%%%%%%%%%%%%%%%

\section{Introduction}

Tidal interactions are known to profoundly impact the orbital evolution of close binary star systems, exoplanetary systems, and moon systems. In binary systems, tides drive the components of the systems toward a synchronized state in which the orbit is circular, and the components have spins that are synchronized and aligned with the orbit. In a compact triple system, no equilibrium state exists, and the endpoint of the orbital evolution is not immediately obvious. Furthermore, the dynamics of tidal interactions coupled with multi-body orbital effects can be quite complex and can lead to the formation of astrophysically interesting systems (e.g., the Jupiter moon system, short-period exoplanetary systems, and compact binary star systems). 

Although there have been many studies of three-body orbital dynamics including tidal dissipation (e.g., Mazeh \& Shaham 1979, Ford et al. 2000, Eggleton \& Kisleva-Eggleton 2001, Fabrycky \& Tremaine 2007, Correia et al. 2011), these studies have primarily treated tides in a parameterized fashion. Some of these studies take resonant orbital effects into account, but their parameterization of tidal interactions ignores resonant tidal effects which may significantly impact the orbital evolution. Furthermore, these studies only consider tidal interactions between two components of the system (i.e., they incorporate tidal interactions between objects 1 and 2, but ignore interactions with object 3). This approximation is justified in most systems because the third body is relatively distant. However, in sufficiently compact triple systems, a more thorough study of three-body tidal effects is necessary. 

Observing tidal effects is very difficult because of the long time scales associated with tidal evolution, and tidal orbital decay has only been observed in a few rare circumstances (such as the orbital decay of a pulsar-MS binary, see Lai 1996,1997; Kumar \& Quataert 1998). Until recently, the direct observation of tidally induced oscillations was difficult because of the extreme precision required ($\sim$ one part in a thousand) over base lines of several days. Fortunately, the continuous observation and high accuracy of the {\it Kepler} satellite is allowing for the direct observation of tidal effects (e.g., Welsh et al. 2011, Thompson et al. 2012) and detailed analyses of tidally excited stellar oscillations (Fuller \& Lai 2012, Burkart et al. 2012). Recently, observations of luminosity variations in the compact triple system HD 181068 (also known as KIC 5952403 or the Trinity system, see Derekas et al. 2011) have provided evidence for three-body tidal effects. To our knowledge, no detailed study of dynamical tides exists for compact triple systems.

We present the first detailed investigation of the tidal excitation of stellar oscillation modes in stars in compact triple systems. Hierarchical three-body systems can experience tidal forcing at frequencies unattainable for two-body systems, allowing for the tidal excitation of high frequency p-modes in the convective envelopes of the stellar components. We investigate the observational signatures of three-body tides, calculate the amplitudes to which stellar oscillation modes are tidally excited, and study the orbital evolution induced by three-body tides.

We also compare our theory to the observations of luminosity fluctuations in HD~181068, accurately characterized by Derekas et al. 2011 and Borkovits et al. 2012. HD~181068 is a triply eclipsing hierarchical star system, with a red giant primary orbited by two dwarf stars. The dwarf stars orbit each other every 0.906 days, and their center of mass orbits the primary every 45.47 days. In addition to eclipses, the lightcurve shows oscillations at combinations of the orbital frequencies. We demonstrate that these oscillations are tidally excited oscillations in the red giant primary excited by the orbital motion of the dwarf stars.

Our paper is organized as follows. In Section \ref{forcing} we derive the strength and frequencies of tidal forcing unique to three-body systems. In Section \ref{modes} we calculate the amplitudes to which modes are tidally excited, and estimate the resulting stellar luminosity variations. In Section \ref{orbit} we calculate the orbital evolution induced by the tidally excited modes. In Section \ref{HD181068}, we describe observations of tidally excited modes in HD~181068, and in Section \ref{comparison} we compare these observations to our theory. In Section \ref{orb} we calculate the possible past and future orbital evolution of systems like HD~181068. Finally, in Section \ref{discussion}, we discuss our results.

\section{three-body Tidal Forcing}
\label{forcing}

\begin{figure*}
\begin{center}
\includegraphics[scale=0.55]{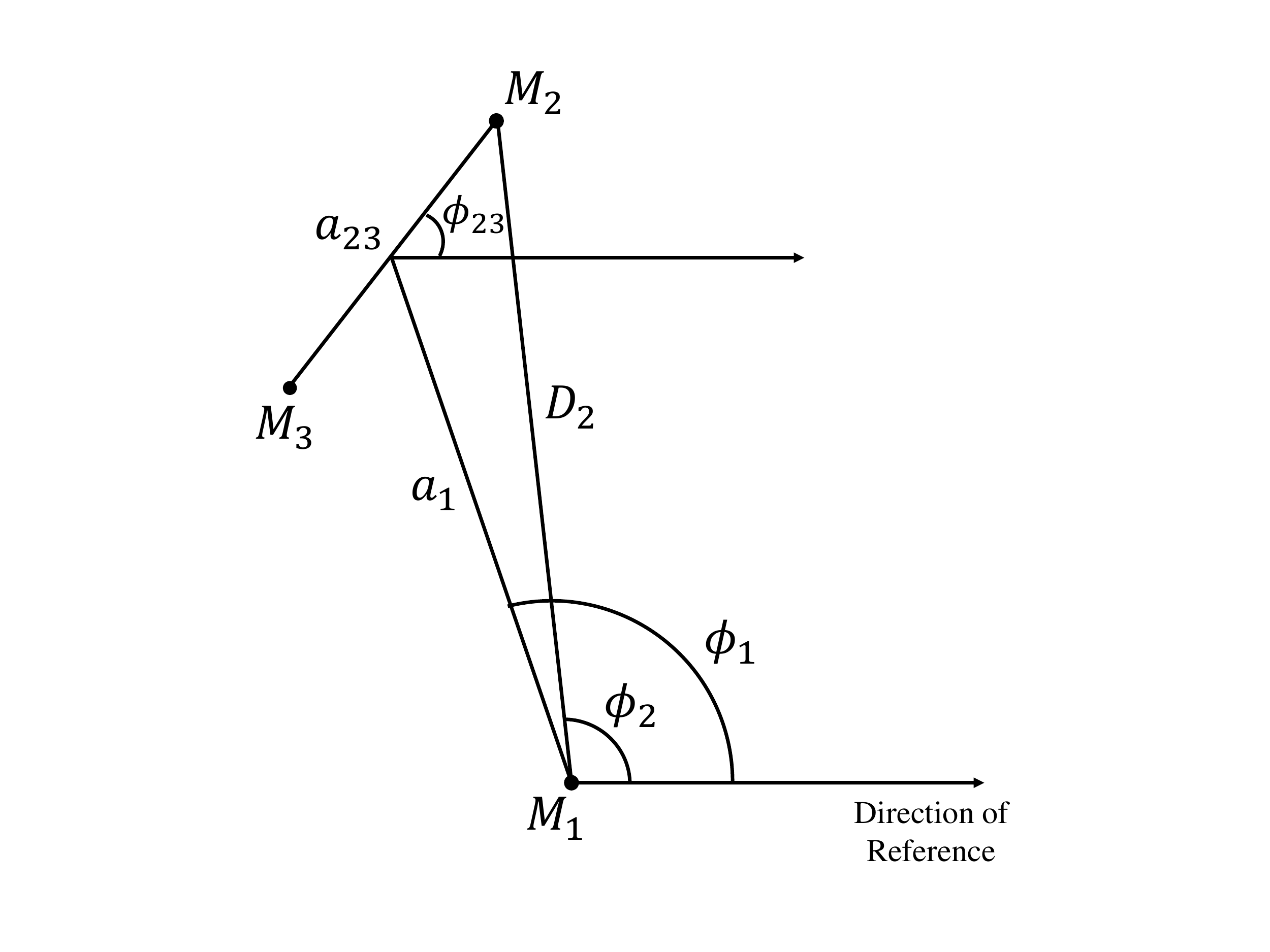}
\end{center}
\caption{ \label{diagram} This diagram (not to scale) depicts the geometry of a triple-star system at arbitrary orbital phase. $a_1$ is the semimajor axis between the center of Star 1 and the center of mass of Stars 2 and 3. $a_{23}$ is the semimajor axis between Stars 2 and 3. $D_2$ is the distance between Stars 1 and 2. $\phi_1$ is the orbital phase (relative to a direction of reference) of the center of mass of Stars 2 and 3, while $\phi_2$ is the orbital phase of Star 2 about Star 1. $\phi_{23}$ is the orbital phase of Stars 2 and 3 about each other relative to the same direction of reference.}
\end{figure*}

Let us consider a triple-star system composed of a central primary star (Star 1), orbited by a pair of companion stars (Stars 2 and 3) at frequency $\Omega_{1}$. Stars 2 and 3 orbit one another at a much higher frequency $\Omega_{23} \gg \Omega_1$. Figure \ref{diagram} shows a diagram of the orbital configuration. We adopt a coordinate system centered on Star 1 with $z-$axis in the direction of its spin angular momentum vector. We consider the case in which all three stars have circular coplanar orbits that are aligned with the stellar spins. We choose a direction of reference such that the observer is located at the azimuthal angle $\phi=0$.

We wish to calculate the form of the time-varying gravitational potential of the short-period binary (Stars 2 and 3) as seen by Star 1. The tidal potential due to Stars 2 and 3 can be decomposed into spherical harmonics as
\be
\label{Ulm}
U_{l,m} = \frac{4\pi}{2l+1}\bigg[\frac{-G M_2}{D_2^{l+1}}r^l Y_{lm}^*({\theta_2},{\phi_2}) + \frac{-G M_3}{D_3^{l+1}}r^l Y_{lm}^*({\theta_3},{\phi_3}) \bigg] Y_{l,m}(\theta,\phi).
\ee
Here, $M_2$, $D_2$,  ${\theta_2}$, and ${\phi_2}$ are the mass, distance, polar angle, and azimuthal angle of Star 2 ($M_3$, $D_3$, ${\theta_3}$, and ${\phi_3}$ are the same quantities for Star 3), while $G$ is the gravitational constant. The dominant terms have $l=|m|=2$ and $l=2$, $m=0$, so we will consider only these terms in our analysis. Since we restrict our analysis to coplanar orbits aligned with the primary spin, $\theta_2=\theta_3=\pi/2$. 

We wish to express the potential in terms of the angular orbital frequencies $\Omega_1$ and $\Omega_{23}$. Some trigonometry reveals that the distance from Star 1 to Star 2 is
\be
D_2^2 = a_1^2 + \Big(\frac{M_3}{M_2+M_3}a_{23}\Big)^2 + 2 \frac{M_3}{M_2+M_3} a_1 a_{23} \cos (\phi_{23}-\phi_1),
\ee
where $\phi_1$ is the phase of the orbit of Stars 2 and 3 about Star 1, $\phi_{23}$ is the phase of the orbit of Star 2 around Star 3, $a_1$ is the semi-major axis between Star 1 and the center of mass of Stars 2 and 3, and $a_{23}$ is the semi-major axis between Stars 2 and 3. Also, the law of sines reveals
\be
\sin (\phi_2-\phi_1) = \frac{M_3}{M_2+M_3} \frac{a_{23}}{D_2} \sin(\phi_{23}-\phi_1).
\ee
In a hierarchical triple system, $a_{23} \ll a_1$, so we expand distances and angles in powers of the small parameter 
\be
\epsilon \equiv a_{23}/a_1.
\ee 
We find
\be
\label{d23}
D_2^{-3} \simeq a_1^{-3} \bigg[1- 3\epsilon \frac{M_3}{M_2+M_3} \cos(\phi_{23}-\phi_1) + \frac{3}{4} \epsilon^2 \bigg(\frac{M_3}{M_2+M_3}\bigg)^2 \bigg(3 + 5\cos\big[2(\phi_{23}-\phi_1)\big]\bigg) \bigg]
\ee
and
\be
\label{phi23}
e^{-i m \phi_2} \simeq e^{-i m \phi_1} \bigg[1 - i m \epsilon \frac{M_3}{M_2+M_3} \sin (\phi_{23}-\phi_1) + \epsilon^2 \bigg(\frac{M_3}{M_2+M_3} \bigg)^2 \bigg( \frac{i m}{2} \sin\big[2(\phi_{23}-\phi_1)\big] - \frac{m^2}{4} + \frac{m^2}{4}\cos\big[2 (\phi_{23}-\phi_1)\big]\bigg)\bigg].
\ee
The values of $D_3^{-3}$ and $e^{-i m \phi_3}$ are the same as equations (\ref{d23}) and (\ref{phi23}), but with 2 and 3 subscripts reversed, and with $\phi_{23} \rightarrow \phi_{23} + \pi$. 

Inserting equations (\ref{d23}) and (\ref{phi23}) and their counterparts for Star 3 into equation (\ref{Ulm}), we find that the zeroth order contribution (the term proportional to $\epsilon^0$) for $l=2$ to the tidal potential is 
\be
\label{U0}
U_{2m}^{(0)} = \frac{-G(M_2+M_3)W_{l,m}r^2}{a_1^3} e^{-im\phi_1} Y_{2,m}(\theta,\phi),
\ee
where $W_{2,0}=-\sqrt{\pi/5}$ and $W_{2,\pm 2}=\sqrt{3\pi/10}$. To zeroth order, the tidal potential due to stars 2 and 3 is simply that of a single star of mass $M_2+M_3$ at a semi-major axis $a_1$. 

Upon summing the contributions from both stars, the first-order terms (proportional to $\epsilon$) in equation ({\ref{Ulm}) vanish. However, the second-order terms produce an additional component of the tidal potential. Dropping terms with no $\phi_1$ or $\phi_{23}$ dependence because they have no time dependence, we find that the second-order component of the tidal potential is
\be
\label{U22}
U_{2,m}^{(2)} = -A_{2,m} r^2 Y_{2,m}(\theta,\phi) \bigg[F_m e^{2i(\phi_{23}-\phi_1)-i m\phi_1} + F_{-m}e^{-2i(\phi_{23}-\phi_1)-i m\phi_1} \bigg],
\ee
where
\be
\label{Alm}
A_{2,m} = \epsilon^2 \frac{G \mu_{23} W_{2,m}}{a_1^{3}},
\ee
$\mu_{23}=M_2 M_3/(M_2+M_3)$, and 
\be
\label{Fm}
F_m = \frac{15+8m+m^2}{8}.
\ee
Then the $m=0$ component yields
\be
\label{U20}
U_{2,0}^{(2)} = -\frac{15}{4} A_{2,0} r^2 P_{2,0}(\cos \theta) \cos\big[2(\phi_{23}-\phi_1)\big].
\ee
We further illuminate the nature of the tidal forcing by adding the $U_{2,2}^{(2)}$ and $U_{2,-2}^{(2)}$ components to find
\be
\label{U222}
U_{2,2}^{(2)} + U_{2,-2}^{(2)}= - A_{2,2} r^2 P_{2,2}(\cos \theta) \bigg[\frac{35}{4} \cos{\Big(2\phi + 2\phi_{23} -4\phi_1\Big)} + \frac{3}{4} \cos{\Big(2\phi - 2\phi_{23}\Big)}\bigg].
\ee
The first term in brackets produces retrograde forcing, i.e., it will excite modes with negative angular momentum that rotate in the opposite direction to the orbital motion of Stars 2 and 3 about one another. The second term in brackets produces prograde forcing, exciting modes with positive angular momentum.

We define $\phi_1=\Omega_1 t$ and $\phi_{23} = \phi_{2,0} + \Omega_{23}t$, where $\phi_{2,0}$ is the phase $\phi_{23}$ of the short-period orbit $t=0$, when their center of mass is closest to the line of sight.  The second-order $m=0$ component of the tidal potential produces forcing at a frequency $\sigma_{0} = 2(\Omega_{23}-\Omega_1)$. The $m=\pm2$ components produce forcing at two frequencies, $\sigma_{-2}=2\Omega_{23}$ and $\sigma_{2}=2(\Omega_{23}-2\Omega_1)$. These frequencies arise because they are the conjunction frequency in a frame corotating with the long-period orbit, but are Doppler shifted by $m\Omega_1$ when viewed in the inertial frame. Due to the hierarchical structure of the system, these frequencies can be larger than the dynamical frequency of Star 1, causing tidal forcing at high frequencies seldom occurring in binary systems.

Equation (\ref{U22}) describes the tidal potential in the inertial frame, but we also wish to calculate the forcing frequencies in the rotating reference frame of Star 1. To do so, we make the transformation $\phi = \phi_r + \Omega_s t$, where $\phi_r$ is the azimuthal angle in the rotating frame and $\Omega_s$ is the angular spin frequency of Star 1. Then in the rotating frame,
\begin{align}
\label{U22r}
U_{2,m,r}^{(2)} &= -A_{2,m} r^2 Y_{2,m}(\theta,\phi_r) \bigg[F_2 e^{2i(\phi_{23}-\phi_1)-i m\phi_1 + i m\Omega_s t} \nonumber \\
&+ F_{-2}e^{-2i(\phi_{23}-\phi_1)-i m\phi_1 + i m\Omega_s t} \bigg].
\end{align}
Thus,
\be
\label{U220r}
U_{2,0,r}^{(2)} = - A_{2,0} r^2 P_{2,0}(\cos \theta) F_0 \cos\big[2(\phi_{23}-\phi_1)\big], 
\ee
and
\begin{align}
\label{U222r}
U_{2,2,r}^{(2)} + U_{2,-2,r}^{(2)} &= -A_{2,2}r^2 P_{2,2}(\cos \theta) \bigg[F_2 \cos{\Big(2\phi_r+2\phi_{23}-4\phi_1 +2\Omega_s t \Big)} \nonumber \\
&+ F_{-2} \cos{\Big(2\phi_r - 2\phi_{23} + 2 \Omega_s t \Big)}\bigg].
\end{align}

Equations (\ref{U22r})-(\ref{U222r}) are general for circular coplanar orbits aligned with the stellar spin. If Star 1 is synchronized with the orbit of Stars 2 and 3 such that $\Omega_{s,1} = \Omega_1$,\footnote{In compact triple systems in which three-body tides operate, it is very likely that $\Omega_{s,1} \simeq \Omega_1$ because the tidal synchronization time due to the zeroth order tidal potential of equation \ref{U0} will be much shorter than tidal time scales due to the three-body tides of equation (\ref{U22}).}
\begin{align}
\label{U222rsynch}
U_{2,2,r}^{(2)} + U_{2,-2,r}^{(2)} &\simeq -A_{2,2} r^2 P_{2,2}(\cos \theta) \bigg[F_2 \cos{\Big(2\phi_r + 2\phi_{23}-2\phi_1 \Big)} \nonumber \\
&+ F_{-2} \cos{\Big(2\phi_r - 2\phi_{23} + 2 \phi_1 \Big)}\bigg].
\end{align}
%\quad {\rm for} \quad \Omega_{s,1|=\Omega_1.
Thus, when the spin of Star 1 is aligned with the orbit of Stars 2 and 3, the absolute values of the forcing frequencies of all three modes (the axisymmetric $m=0$ mode, and the prograde and retrograde $m=\pm2$ azimuthal modes) in the corotating frame are identical, namely, $|\nu|=2(\Omega_{23}-\Omega_1)$.

\section{Mode Excitation and Observation}
\label{modes}

\subsection{Mode Amplitudes}
\label{amp}

With the tidal potential known, we may calculate the amplitude of the tidally forced modes. In this section we consider only tidal forcing by the $l=2$ second-order component of the tidal potential. We use the subscript $\alpha$ to refer to a mode of angular degree $l$ and $m$, with a frequency in the rotating frame of $\omega_\alpha$. In this frame, the mode amplitude $c_\alpha$ satisfies (Schenk et al. 2001)
\begin{align}
\label{c}
\dot{c}_\alpha +(i \omega_\alpha + \gamma_\alpha)c_\alpha &= \frac{i}{2\varepsilon_\alpha}\langle\bxi_\alpha(\br),-\nabla U \rangle \nonumber \\
&=\frac{i A_{\alpha} Q_\alpha}{2 \varepsilon_\alpha} e^{im(\Omega_s  -\Omega_1)t} \bigg[F_m e^{2i(\Omega_{23}-\Omega_1)t + 2i\phi_{2,0}} + F_{-m} e^{-2i(\Omega_{23}-\Omega_1)t-2i\phi_{2,0}} \bigg].
\end{align}
Here, $\omega_\alpha = \varepsilon_\alpha - m C_{\alpha} \Omega_s {\rm sgn}(\varepsilon_\alpha)$ is the mode frequency in the rotating frame, $\varepsilon_\alpha$ is the unperturbed frequency for a non-rotating star, $C_{\alpha}$ is the rotational kernel (see Fuller \& Lai 2012), and $\gamma_\alpha$ is the mode damping rate. In this formalism, both $m$ and $\varepsilon_\alpha$ can have positive or negative values. The dimensionless coefficient $Q_\alpha$ is the overlap integral of the mode with the tidal potential, defined as
\begin{align}
Q_\alpha &= \frac{1}{M_1R_1^l}\langle \bxi_\alpha | \nabla (r^l Y_{lm}) \rangle \nonumber \\
 &= \frac{1}{M_1R_1^l}\int\!d^3x\,\,\rho\bxi_\alpha^\ast\cdot \nabla (r^lY_{lm}),
\end{align}
where $\bxi_\alpha$ is the mode eigenvector normalized via the condition 
\be
\label{norm}
\langle \bxi_\alpha | \bxi_\alpha \rangle = M_1 R_1^l,
\ee
and $R_1$ is the radius of Star 1.

The non-homogeneous solution of equation (\ref{c}) is
\be
\label{c2}
c_{\alpha} = \frac{A_{\alpha} Q_\alpha}{2 \varepsilon_\alpha} \bigg[\frac{F_m e^{i\nu_m t + 2i\phi_{2,0}}}{\omega_\alpha + \nu_m - i\gamma_\alpha} + \frac{F_{-m} e^{-i\nu_{-m} t- 2i\phi_{2,0}}}{\omega_\alpha - \nu_{-m} - i\gamma_\alpha}\bigg]
\ee
where 
\be
\label{omegaf}
\nu_m=2(\Omega_{23} - \Omega_1) + m(\Omega_s -\Omega_1).
\ee
The total tidal response of the star is $\bxi({\bf r},t) = \sum_\alpha c_\alpha(t) \bxi_\alpha({\bf r})$. We assume that the star is slowly rotating ($\Omega_s \ll \omega_\alpha$) such that $\xi_\alpha(r)$ and $Q_\alpha$ are independent of $m$ or $\Omega_s$, and $\omega_\alpha \simeq \varepsilon_\alpha$. Summing over both signs of $\omega_\alpha$ yields 
\be
\label{bxi}
\bxi({\bf r},t) = \sum_{\alpha,\omega > 0} A_\alpha Q_\alpha \bxi_\alpha({\bf r}) \bigg[\frac{F_m e^{i \nu_m t + 2i\phi_{2,0}}}{\omega_{\alpha}^2 - \nu_m^2 + 2 i \nu_m \gamma_{\alpha} + \gamma_\alpha^2} + \frac{F_{-m} e^{-i \nu_{-m} t - 2i\phi_{2,0}}}{\omega_{\alpha}^2 - \nu_{-m}^2 - 2 i \nu_{-m} \gamma_{\alpha} + \gamma_\alpha^2}\bigg].
\ee
Summing over both signs of $m$ then yields
\begin{align}
\label{bxitotm0}
\bxi({\bf r},t,m=0) &= \sum_{\alpha,\omega > 0, m=0} 2 A_\alpha Q_\alpha P_{2,m}(\cos \theta) \bxi_\alpha(r) F_m D_{\alpha,m} \cos \big(\nu_m t + m \phi_r + 2\phi_{2,0} + \psi_{\alpha,m} \big),
\end{align}
\begin{align}
\label{bxitotm2}
\bxi({\bf r},t,m=\pm2) &= \sum_{\alpha,\omega > 0, m=2} 2 A_\alpha Q_\alpha P_{2,m}(\cos \theta) \bxi_\alpha(r) \bigg[F_m D_{\alpha,m} \cos \big(\nu_m t + m \phi_r + 2\phi_{2,0} + \psi_{\alpha,m} \big) \nonumber \\
&+ F_{-m} D_{\alpha,-m} \cos \big(\nu_{-m} t - m \phi_r + 2\phi_{2,0} + \psi_{\alpha,-m} \big)  \bigg],
\end{align}
with 
\be
\label{D}
D_{\alpha,m} = \Big[\big(\omega_\alpha^2 - \nu_m^2 \big)^2 + 4 \nu_m^2\gamma_\alpha^2 + \gamma_\alpha^4 \Big]^{-1/2}
\ee
and
\be
\label{psi}
\psi_{\alpha,m} = \frac{\pi}{2} + \tan^{-1} \bigg(\frac{\omega_\alpha^2 - \nu_m^2 + \gamma_\alpha^2}{2\nu_m\gamma_\alpha}\bigg).
\ee

The $m=0$ modes produce a radial displacement 
\be
\label{bxim0}
\xi_r({\bf r},t,m=0) = \sum_{\alpha,\omega > 0,m=0} 2 A_{\alpha} Q_\alpha F_m D_{\alpha,m} \xi_{r,\alpha}(r) P_{2,0}(\cos \theta) \cos(\nu_m t + 2\phi_{2,0} + \psi_{\alpha,m}).
\ee
The $m=\pm2$ modes produce a radial displacement
\begin{align}
\label{bxim2}
\xi_r({\bf r},t,m=\pm2) &\simeq \sum_{\alpha,\omega > 0,m=2} 2 A_{\alpha} Q_\alpha F_m D_{\alpha,m} \xi_{r,\alpha}(r) P_{2,2}(\cos \theta) \cos(\nu_m t + 2 \phi_r + 2\phi_{2,0} +\psi_{\alpha,m}) \nonumber \\
&+ 2 A_{\alpha} Q_\alpha F_{-m} D_{\alpha-,m} \xi_{r,\alpha}(r) P_{2,2} (\cos\theta) \cos(\nu_{-m} t - 2 \phi_r + 2\phi_{2,0} +\psi_{\alpha,-m}).
\end{align}
The first term is the radial displacement due to the retrograde mode, and the second is due to the prograde mode.

\subsection{Luminosity Variations}

The luminosity variation produced by a mode, $\Delta L_\alpha/L$, is not trivial to calculate, because the temperature fluctuation produced by the mode is sensitive to non-adiabatic effects and subtleties in the outer boundary condition. We refer the reader to other studies (e.g., Buta \& Smith 1979, Robinson et al. 1982) which have attempted to quantify the visibility of a mode given its amplitude and eigenfunction. If the mode is adiabatic, its luminosity fluctuation is 
\be
\label{deltaL}
\frac{\Delta L_\alpha}{L} = 2 A_{\alpha} Q_\alpha F_m D_{\alpha,m} \xi_{r,\alpha}(R) V_\alpha Y_{l,m}(\theta_o,\phi_o) \cos(\sigma_m t + 2\phi_{2,0} + \psi_{\alpha,m})
\ee
with
\be
\label{sig}
\sigma_m=2(\Omega_{23}-\Omega_1)-m\Omega_1,
\ee
and here $m=2$ corresponds to the retrograde mode and $m=-2$ corresponds to the prograde mode. The angular coordinates $\theta_o$ and $\phi_o$ indicate the direction of the observer relative to the symmetry axis of the mode, here assumed to be the spin axis of Star 1. We define $V_\alpha$ to be the visibility function of the mode, which is dependent on geometrical, opacity, and non-adiabatic effects. In high-inclination systems like HD 181068, $\theta_o \simeq 90^\circ$, and we have chosen our coordinate system such that $\phi_o=0$. In this case, $Y_{l,m}(\theta_o,\phi_o) \simeq -\sqrt{5/(16\pi)}$ for $m=0$ modes and $Y_{l,m}(\theta_o,\phi_o) \simeq \sqrt{15/(32\pi)}$ for $m=\pm2$ modes. 

The value of $V_\alpha$ is uncertain. If non-adiabatic effects are significant, $V_\alpha$ will be complex and the luminosity variation will be phase-shifted from equation (\ref{deltaL}). Here we consider adiabatic modes, but we remember that an observation of a phase shift is indicative of non-adiabatic effects. In the adiabatic limit, Buta \& Smith (1979) find
\be
\label{V2}
V_\alpha = \alpha_l + \beta_l + 4 \nabla_{\rm ad} \gamma_l H_\alpha,
\ee
where $\alpha_l$, $\beta_l$, and $\gamma_l$ are coefficients of order unity that depend on the spherical harmonic $l$ of the mode and the stellar limb-darkening function (we use the values given in Buta \& Smith 1979), and $\nabla_{\rm ad}$ is the adiabatic temperature gradient at the surface. The first term in equation (\ref{V2}) is due to surface area distortions, the second is due to surface normal distortions, and the third term is due to temperature effects. The function $H_\alpha$ describes the magnitude of the pressure perturbation compared to the radial surface displacement, i.e., $\Delta T_\alpha/T = H_\alpha \Delta P_\alpha/P$.  According to Dziembowski (1971), the value of $H_\alpha$ for adiabatic oscillations is
\be
\label{F}
H_\alpha = \bigg[\frac{l(l+1)}{\bar{\omega}_\alpha^2} -4 - \bar{\omega}_\alpha^2\bigg],
\ee
where
\be
\label{baromega}
\bar{\omega}_\alpha = \frac{\omega_\alpha}{\omega_{\rm dyn}},
\ee
and $\omega_{\rm dyn} = \sqrt{GM_1/R_1^3}$ is the dynamical frequency of Star 1. In this description, $H_\alpha$ is large for high-order g-modes (low values of $\bar{\omega}_\alpha$) or high-order p-modes (high values of $\bar{\omega}_\alpha$).

Gouttebroze \& Toutain (1994) have attempted to estimate the value of $V_\alpha$ for p-modes modes in a solar model. They calculated non-adiabatic mode eigenfunctions, and then calculated luminosity variations by adding up the perturbed flux from a grid of emitting surface elements. They found that $V_\alpha$ is of order unity for low-order p-modes, but has substantially larger values for the high-order p-modes that typically produce solar-like oscillations. In the analysis of Section \ref{comparison}, we calculate $V_\alpha$ from equation (\ref{V2}), considering a lower limit of $H_\alpha \approx 1$ and an upper limit of equation (\ref{F}).

\section{Effect of Modes on Orbital Evolution}
\label{orbit}

\subsection{Hamiltonian Formalism}

The tidally excited modes draw energy and angular momentum from the orbital motions and deposit them in Star 1. The effect of the modes on the orbits can be calculated from the Hamiltonian of the gravitational interaction between the modes and the stars. This Hamiltonian is
\be
\label{H}
H = \int d^3 r U({\bf r},t) \sum_{\alpha} c^*_{\alpha}(t) \delta \rho^{*}_\alpha({\bf r}), 
\ee
where $\delta \rho$ is the Eulerian density perturbation associated with each mode. Performing the integration over the volume of Star 1, and considering only $l=2$ terms yields
\be
\label{H2}
H = -M_1 R_1^2 \sum_{\alpha, \omega>0} A_\alpha Q_\alpha \bigg[F_m e^{2i(\phi_{23}-\phi_1)-i m\phi_1 + i m\Omega_s t} + F_{-m}e^{-2i(\phi_{23}-\phi_1)-i m\phi_1 + i m\Omega_s t} \bigg] c^*_{\alpha}(t). 
\ee
Inserting equation (\ref{c2}) into equation (\ref{H2}), we find
\begin{align}
\label{H3}
H &= -M_1 R_1^2 \sum_{\alpha, \omega>0} (A_\alpha Q_\alpha)^2 \frac{\omega_\alpha}{\varepsilon_\alpha} \bigg[F_m e^{2i(\phi_{23}-\phi_1)-i m\phi_1} + F_{-m}e^{-2i(\phi_{23}-\phi_1)-i m\phi_1} \bigg] \nonumber \\
& \times \bigg[\frac{F_m e^{-i\sigma_{f1} t}}{\omega_\alpha^2 - \nu_m^2 - 2 i \nu_m \gamma_{\alpha} + \gamma_\alpha^2} + \frac{F_{-m} e^{-i\sigma_{f2} t}}{\omega_\alpha^2 - \nu_{-m}^2 + 2 i \nu_{-m} \gamma_{\alpha} + \gamma_\alpha^2}\bigg], 
\end{align}
with $\sigma_{f1}=2(\Omega_{23} - \Omega_1) - m\Omega_1$, and $\sigma_{f2}=-2(\Omega_{23} - \Omega_1) -m\Omega_1$. 

The orbital evolution equations are $\dot{L}_1 = -  d H/d \phi_1$, $\dot{L}_{23} = -  d H/d \phi_{23}$, where $L_1 = \mu_1 a_1^2 \Omega_1$ and $L_{23} = \mu_{23} a_{23}^2 \Omega_{23}$ are the angular momenta of the outer and inner orbit, respectively, and $\mu_1 = M_1(M_2+M_3)/(M_1+M_2+M_3)$. In the limit $\Omega_s \ll \omega_\alpha$ we find
\begin{align}
\label{dL1dt}
\dot{L}_1 &= M_1 R_1^2 \sum_{\alpha, \omega>0} (A_\alpha Q_\alpha)^2 i \bigg[\frac{-(2+m)F_m^2}{\omega_\alpha^2 - \nu_m^2 - 2 i \gamma_\alpha \nu_m + \gamma_\alpha^2} \nonumber \\
&+  \frac{(2-m)F_{-m}^2}{\omega_\alpha^2 - \nu_{-m}^2 + 2 i \gamma_\alpha \nu_{-m} + \gamma_\alpha^2} \bigg]
\end{align}
and
\begin{align}
\label{dL23dt}
\dot{L}_{23} &= M_1 R_1^2 \sum_{\alpha, \omega>0} (A_\alpha Q_\alpha)^2 i \bigg[\frac{2F_m^2}{\omega_\alpha^2 - \nu_m^2 - 2 i \gamma_\alpha \nu_m + \gamma_\alpha^2} \nonumber \\
&+  \frac{-2F_{-m}^2}{\omega_\alpha^2 - \nu_{-m}^2 + 2 i \gamma_\alpha \nu_{-m} + \gamma_\alpha^2} \bigg].
\end{align}
We have discarded rapidly oscillating terms because they produce no secular variations. The torque due to the $m=0$ modes is
\be
\label{dL1dtm0}
\dot{L}_1(m=0) = 8 M_1 R_1^2 \sum_{\alpha, \omega>0,m=0} (A_\alpha Q_\alpha)^2  F_m^2 \Gamma_{\alpha,m}, 
\ee
where
\be
\label{biggamma}
\Gamma_{\alpha,m} = \frac{\nu_m \gamma_\alpha}{{(\omega_\alpha^2 - \nu_m^2)^2 + 4(\nu_m \gamma_\alpha)^2 + \gamma_\alpha^4}},
\ee
and $\dot{L}_{23}(m=0) = - \dot{L}_1(m=0)$. The torque due to $m=\pm2$ modes is
\be
\label{dL1dtm2}
\dot{L}_1(m=\pm 2) = 4 M_1 R_1^2 \sum_{\alpha, \omega>0,m=2} (A_\alpha Q_\alpha)^2 (2+m) F_m^2 \Gamma_{\alpha,m}
\ee
and
\be
\label{dL23dtm2}
\dot{L}_{23}(m=\pm 2) = 8 M_1 R_1^2 \sum_{\alpha, \omega>0,m=2} (A_\alpha Q_\alpha)^2 \bigg[ F_m^2 \Gamma_{\alpha,m} + F_{-m}^2 \Gamma_{\alpha,-m} \bigg].
\ee

Equations (\ref{dL1dtm0}-\ref{dL23dtm2}) may be written in terms of a dimensionless damping rate $\bar{\Gamma}_{\alpha,m} = \Gamma_{\alpha,m} \omega^2_{\rm dyn}$, such that we obtain the more familiar scaling for tidal dissipation:
\be
\label{dL1dtm0b}
\dot{L}_1(m=0) = 8 \frac{G M_1^2}{R_1} \epsilon^2 \bigg(\frac{\mu_{23}}{M_1}\bigg)^2 \bigg(\frac{R_1}{a_1}\bigg)^6  \sum_{\alpha, \omega>0} (W_\alpha Q_\alpha F_0)^2 \bar{\Gamma}_{\alpha,0}, \quad {\rm for} \quad m=0 \quad {\rm modes}, 
\ee
\be
\label{dL1dtm2b}
\dot{L}_1(m=\pm 2) = 16 \frac{G M_1^2}{R_1} \epsilon^2 \bigg(\frac{\mu_{23}}{M_1}\bigg)^2 \bigg(\frac{R_1}{a_1}\bigg)^6  \sum_{\alpha, \omega>0} (W_\alpha Q_\alpha F_2)^2 \bar{\Gamma}_{\alpha,2}, \quad {\rm for} \quad m=\pm 2 \quad {\rm modes}, 
\ee
\be
\label{dL23dtm0b}
\dot{L}_{23}(m=0) = -8 \frac{G M_1^2}{R_1} \epsilon^2 \bigg(\frac{\mu_{23}}{M_1}\bigg)^2 \bigg(\frac{R_1}{a_1}\bigg)^6  \sum_{\alpha, \omega>0} (W_\alpha Q_\alpha F_0)^2 \bar{\Gamma}_{\alpha,0}, \quad {\rm for}\quad  m=0 \quad {\rm modes}. 
\ee
\be
\label{dL23dtm2b}
\dot{L}_{23}(m=\pm 2) = -8 \frac{G M_1^2}{R_1} \epsilon^2 \bigg(\frac{\mu_{23}}{M_1}\bigg)^2 \bigg(\frac{R_1}{a_1}\bigg)^6  \sum_{\alpha,\omega>0} (W_\alpha Q_\alpha)^2 \bigg[ F_2^2 \bar{\Gamma}_{\alpha,2} + F_{-2}^2\bar{\Gamma}_{\alpha,-2}\bigg], \quad {\rm for}\quad  m=\pm 2 \quad {\rm modes}. 
\ee
Conservation of angular momentum requires that the torque on Star 1 is $\dot{L}_* = -\dot{L}_1 - \dot{L}_{23}$, or
\be
\label{dLstardt}
\dot{L}_* = -8 \frac{G M_1^2}{R_1} \epsilon^2 \bigg(\frac{\mu_{23}}{M_1}\bigg)^2 \bigg(\frac{R_1}{a_1}\bigg)^6 \sum_{\alpha, \omega>0} (W_\alpha Q_\alpha)^2 \bigg[ F_2^2 \bar{\Gamma}_{\alpha,2} - F_{-2}^2\bar{\Gamma}_{\alpha,-2}\bigg].
\ee

We can see that $m=0$ modes draw angular momentum from the inner orbit and transfer it to the outer orbit. The $m=2$ (retrograde) modes take angular momentum from the inner orbit and the spin of Star 1, depositing it in the outer orbit. The $m=-2$ (prograde) modes draw angular momentum from the inner orbit and transfer it to the spin of Star 1. Thus, in all cases, three-body tides cause the inner orbit to decay and the outer orbit to expand, although the outer orbit expands by a much smaller factor because of its larger moment of inertia. Star 1 can be either spun up or spun down, depending on which mode (prograde or retrograde) contains more energy. Under most circumstances, Star 1 will usually be spun down because $F_2 > F_{-2}$. In the limit of zero eccentricity, the orbital energies change as $\dot{E}_1 = \Omega_1 \dot{L}_1$ and $\dot{E}_{23} = \Omega_{23} \dot{L}_{23}$. Consequently, $\dot{e}_1 = \dot{e}_{23} = 0$, and the orbits remain circular. Finally, we note that three-body tides tend to increase the stability of hierarchical triples because they cause the period ratio $P_1/P_{23}$ to increase.

%\footnote{Equation (\ref{dLstardt}) can also be derived by calculating the rate at which modes deposit angular momenta in Star 1: $dL_*/dt = M_1 R_1^2 \sum_\alpha m \gamma_\alpha \omega_{f} |c_\alpha|^2$.}

\section{Observations of HD~181068}
\label{HD181068}

\begin{table*}
\begin{center}
\caption{Properties of the HD 181068 system as measured by Borkovits et al. 2012.}
\begin{tabular}{@{}lccc}
\hline\hline
$ \quad \quad $ & $M (M_\odot)$ & $R (R_\odot)$ & $T_{\rm eff} (K)$ \\
\hline
Star 1 & $3.0 \pm 0.1$ & $12.46 \pm 0.15$ & $5100 \pm 100$ \\
\hline
Star 2 & $0.915 \pm 0.034$ & $0.865 \pm 0.010$ & $5100 \pm 100$ \\
\hline
Star 3 & $0.870 \pm 0.043 $ & $0.800 \pm 0.020$ & $4675 \pm 100$ \\
\hline\hline
\end{tabular}
\end{center}
\end{table*}

We now apply our theory to the compact triple-star system HD 181068, also known as the Trinity system, whose properties are listed in Table 1 (see Derekas et al. 2011, Borkovits et al. 2012). Stars 2 and 3 orbit about Star 1 at an angular frequency $\Omega_1 = 0.138 \ {\rm d}^{-1}$, while Stars 2 and 3 orbit about each other at $\Omega_{23} = 6.94 \ {\rm d}^{-1}$. All three stars are nearly exactly coplanar, with orbital inclinations of $i\simeq90^\circ$. The red giant primary has a radius and surface temperature typical of red clump stars, with $R_1 = 12.46 \pm 0.15 R_\odot$ and $T_{\rm eff} = 5100 \pm 100$K, but has a fairly large mass of $M_1 = 3.0 \pm 0.1 M_\odot$.

In the discovery paper of HD181068 (Derekas et al. 2011), 218 days of {\it Kepler} data were analyzed. One of the most surprising results was that the main component of the system, which is a red giant star, did not show solar-like oscillations. Instead, other pulsations were detected with frequencies close to double the orbital frequency of the short-period binary. Derekas et al. 2011 suggested that these pulsations might be tidally forced oscillations and that there might be a mechanism which suppresses the solar-like oscillations.

\subsection{Fourier Analysis}

As of writing, 11 quarters of data from Kepler have been made available to us, representing almost three years of essentially uninterrupted observations. The first six quarters of data were obtained in long-cadence mode (one point every 29.4 minutes), while Q7 to Q11 were taken in short-cadence mode (one point every 58.9 seconds). Here we present a period analysis of the combined long-cadence and short-cadence data, the latter covering 450 days.

As a first step, we subtracted the eclipses and rotational variations by using the light curve fit of Borkovits et al. 2012, which resulted in a nearly continuous data set containing the pulsations. For the period analysis, we used Period04 by Lenz \& Breger 2005. In the Fourier-spectra, a number of the peaks were located ${\rm < 0.1 d^{-1}}$ indicating long term  variability, remnants of the light curve fit, or instrumental effects that are negligible in the present analysis. Peaks with significance are listed in Table\ \ref{fourier} and the Fourier-spectrum is shown in Figure \ref{fourierlc}.

\begin{table}
\begin{center}
 \caption{\label{fourier} The significant peaks of the period analysis for HD~181068. A number of peaks located ${\rm < 0.1 d^{-1}}$ were left out of the analysis.}   
\label{rylepfoupar}   
\begin{tabular}{|cccccc|}   
\hline\hline
No. & Frequency & Orbital Relation & Amplitude & Phase & S/N \\  
& (${\rm d^{-1}}$) & & (mmag) &   \\
\hline  
$f_{1}$ & 2.1203 & $2(\Omega_{23} - 2 \Omega_1)/(2\pi)$ & 0.44 & 0.0403 & 58  \\
$f_{2}$ & 2.1643 & $2(\Omega_{23} - \Omega_1)/(2\pi)$ & 0.25 & 0.2889 & 35  \\
$f_{3}$ & 1.1065 & $\Omega_{23}/(2\pi)$ & 0.08 & 0.2843 & 3.4  \\
$f_{4}$ & 2.2084 & $2\Omega_{23}/(2\pi)$ & 0.048 & 0.8719 & 6.8  \\
\hline\hline   
\end{tabular} 
\end{center}  
\end{table}

The most intriguing result of the period analysis is that $f_{1}$ and $f_{2}$ are linear combinations of the two orbital frequencies, suggesting a tidal origin. Their angular frequencies are separated by $2\Omega_1$, creating a beat pattern in which they are in phase near the primary eclipses and occultations (see Figure \ref{beating}). The peaks frequencies $f_3$ and $f_4$ correspond to one and two times the orbital frequency of Stars 2 and 3, and may be caused by the imperfect subtraction of the eclipses or spots on the components. However, they could have a tidal component as well. 

\begin{figure}
\begin{center}
\includegraphics[width=8cm]{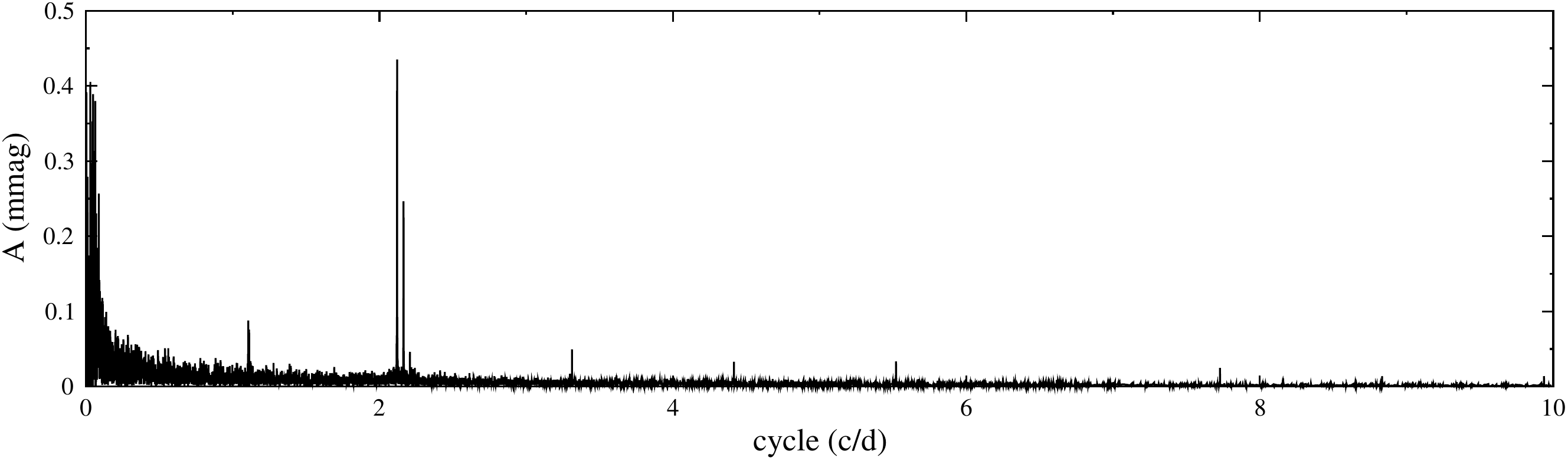}
\end{center}
\caption{\label{fourierlc} Fourier spectrum of 11 quarters of long-cadence data.}
\end{figure}

\begin{figure}
\begin{center}
\includegraphics[width=8cm]{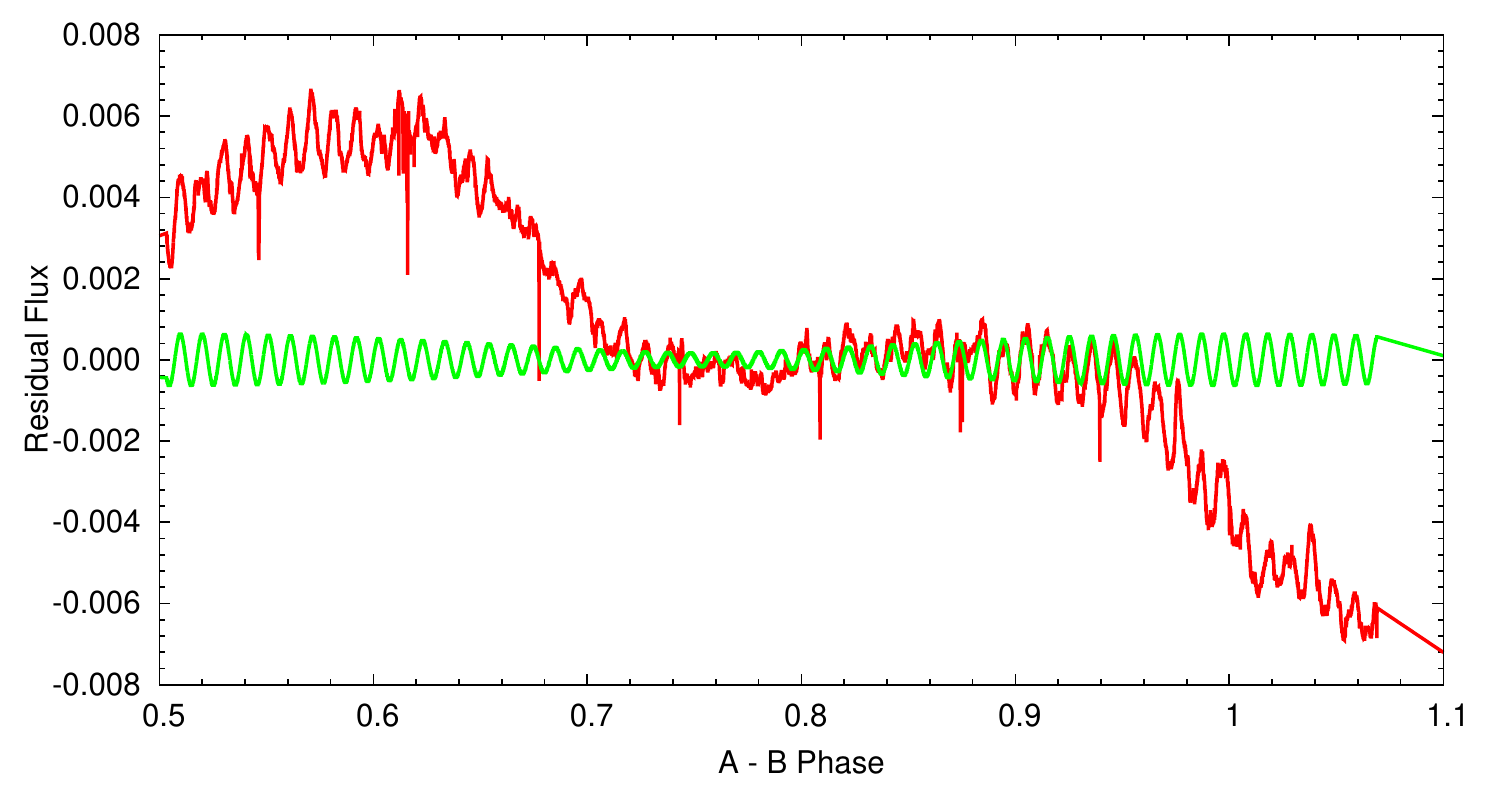}
\end{center}
\caption{\label{beating} {\it Red curve:} Sample of light curve of HD~181068 between long duration eclipses. The $x-$axis is the phase of the long-period orbit, measured from the primary minimum at BJD~55545.466. {\it Green curve:} The simulated light curve of the oscillations. Note the beating pattern due to two close frequency oscillations.}
\end{figure}

\subsection{Lack of solar-like oscillations}

Red giant stars are well known to show solar-like oscillations (De Ridder et al. 2009, Chaplin et al. 2011a) which are excited by near-surface convection (see, e.g., Houdek et al. 1999, Samadi et al. 2007). Nearly all red giants observed by Kepler show detectable oscillations (Huber et al. 1010, Kallinger et al. 2010, Hekker et al. 2011), which can be used to probe the internal properties of the star (e.g., Bedding et al. 2011, Beck et al. 2012, Mosser et al. 2012). The frequency of maximum power (\numax) of solar-like oscillations can be estimated using the scaling relation (Brown et al. 1991)
\begin{equation}
\nu_{\rm max} = \frac{ M/M_{\sun}(T_{\rm eff}/T_{\rm eff,\sun})^{3.5}}{L/L_{\sun}} \nu_{\rm max,\sun} ,
\label{equ:nmax}
\end{equation}
with $\nu_{\rm max,\sun} = 3090 \muHz$. Using the fundamental properties given by Borkovits et al. 2012, we calculate $\numax = 64\pm7\muHz$ ($5.5\pm0.6$ c/d) for the red giant component. To search for solar-like oscillations in HD181068, we first remove all primary and secondary eclipses from the data using the ephemeris given by Borkovits et al. 2012. To further remove long-periodic variability we then apply a Savitzky-Golay filter (Savitzky \& Golay 1964) with a width of 5 days. Finally, we pre-whiten the most significant low frequency variations due to tidal oscillations, as discussed in the previous section.

The upper panel of Figure \ref{fig:ps} shows a power spectrum of the residual Q1-Q11 long-cadence light curve. For comparison the middle panel shows the power spectrum of KIC4662939, a Helium-core burning red giant with similar fundamental properties as red giant component in HD181068 (Bedding et al. 2011). Despite an increase of the data set length by a factor of three compared to Derekas et al. 2011, the lack of solar-like oscillations in HD181068\,A is clearly confirmed. On the other hand, we observe that both stars exhibit a similar decrease of power from low to high frequencies, which is the typical signature of granulation (see Mathur et al. 2011). This confirms that both stars indeed have similar fundamental properties, but that solar-like oscillations are suppressed in HD181068\,A.

We speculate that the lack of solar-like oscillations of the main component is related to the close multiplicity of the components of HD 181068. Derekas et al. 2011 measured a rotational velocity of Star 1 of $v \sin i = 14 \pm 1 \ {\rm km \ s}^{-1}$, corresponding to $\Omega_{s,1} = 0.14 \pm 0.01 \ {\rm d}^{-1}$. This spin frequency is consistent with Star 1 being synchronized with the long-period orbit, indicating that it has been tidally synchronized. It is also an abnormally large spin frequency for red giants (de Medeiros et al. 1996), and the rapid rotation may generate a strong magnetic dynamo. Indeed, the light curve of HD 181068 exhibits flaring events (some events occur during occultations of Stars 2 and 3, indicating the flares originate from Star 1, see Borkovits et al. 2012) that indicate high levels of magnetic activity. Chaplin et al. 2011b has shown that solar-like oscillations are suppressed in abnormally active stars, presumably due to their rapid rotation or due to the effects of strong magnetic fields. We therefore speculate that the tidal synchronization of the primary in HD 181068 creates rapid rotation and high magnetic activity that suppress the excitation of solar-like oscillations.

\begin{figure}
\begin{center}
\resizebox{12cm}{!}{\includegraphics{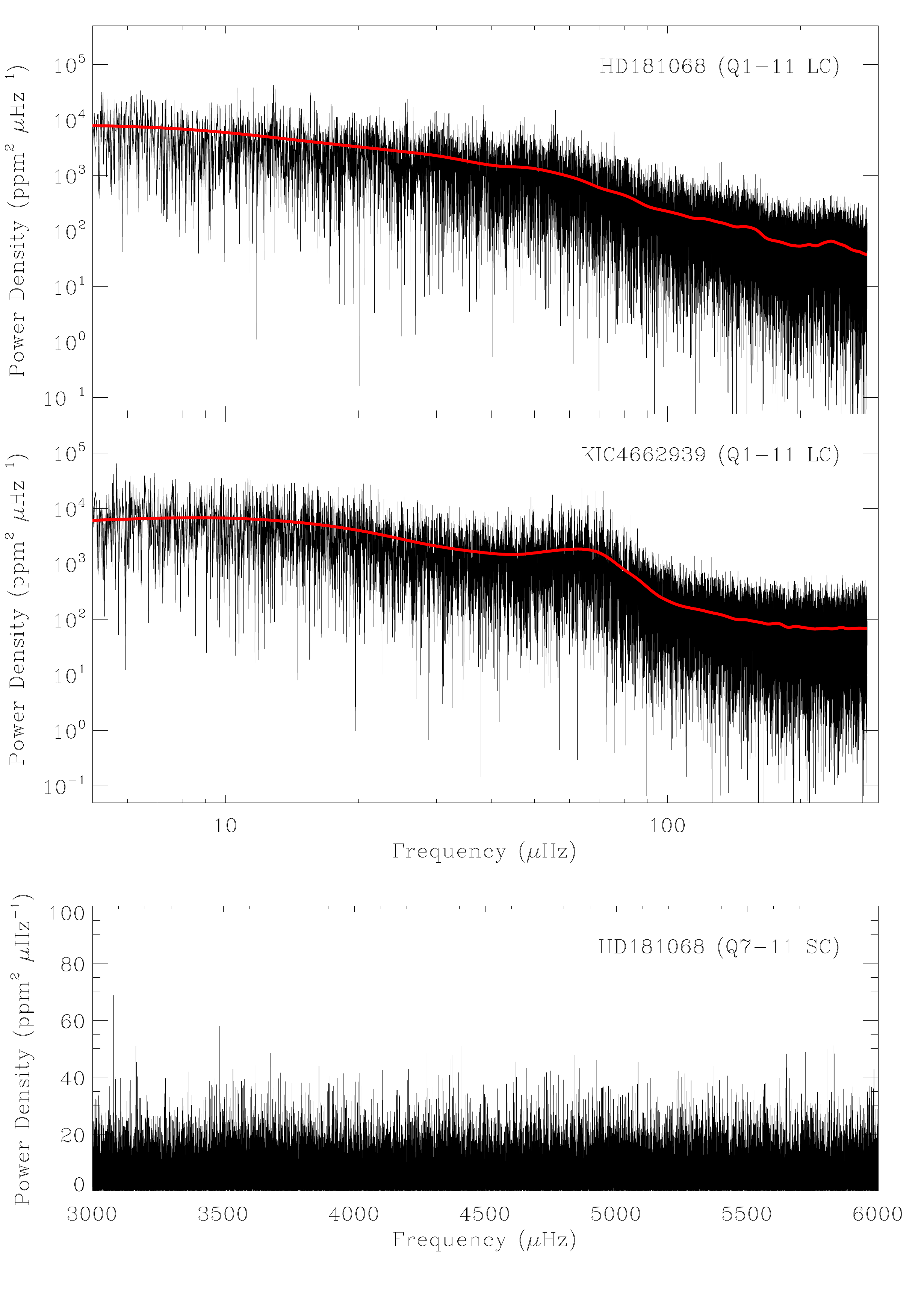}}
\caption{Top panel: Power spectrum of the Q1-Q11 long-cadence data of HD~181068 after removing eclipses and low frequency variability. The red line shows the power spectrum smoothed with a Gaussian filter with a full-width at half maximum of 5\muHz. Middle panel: Same as top panel but for KIC~4662939. Bottom panel: Power spectrum of the Q7-11 short-cadence data of HD~181068 after removing eclipses and low frequency variability.}
\label{fig:ps} 
\end{center}
\end{figure}

The available short-cadence data also allow us to search for solar-like oscillations in the dwarf components. Using equation (\ref{equ:nmax}), we expect the dwarf components to oscillate at frequencies between $3500-5000\muHz$. The bottom panel of Figure \ref{fig:ps} shows the power spectrum of the short-cadence data after performing the same corrections as described above. We do not detect any significant power excess in the data. Given the much lower luminosities of the dwarf companions and the amplitude dilution by the  brighter giant component, this implies that the amplitudes are too small to be detected with the data at hand. Solar-like oscillations in the dwarfs may be also suppressed by the same mechanisms described above.

\section{Comparison With Observations}
\label{comparison}

\subsection{Stellar Model}
\label{model}

\begin{figure*}
\begin{center}
\includegraphics[scale=0.5]{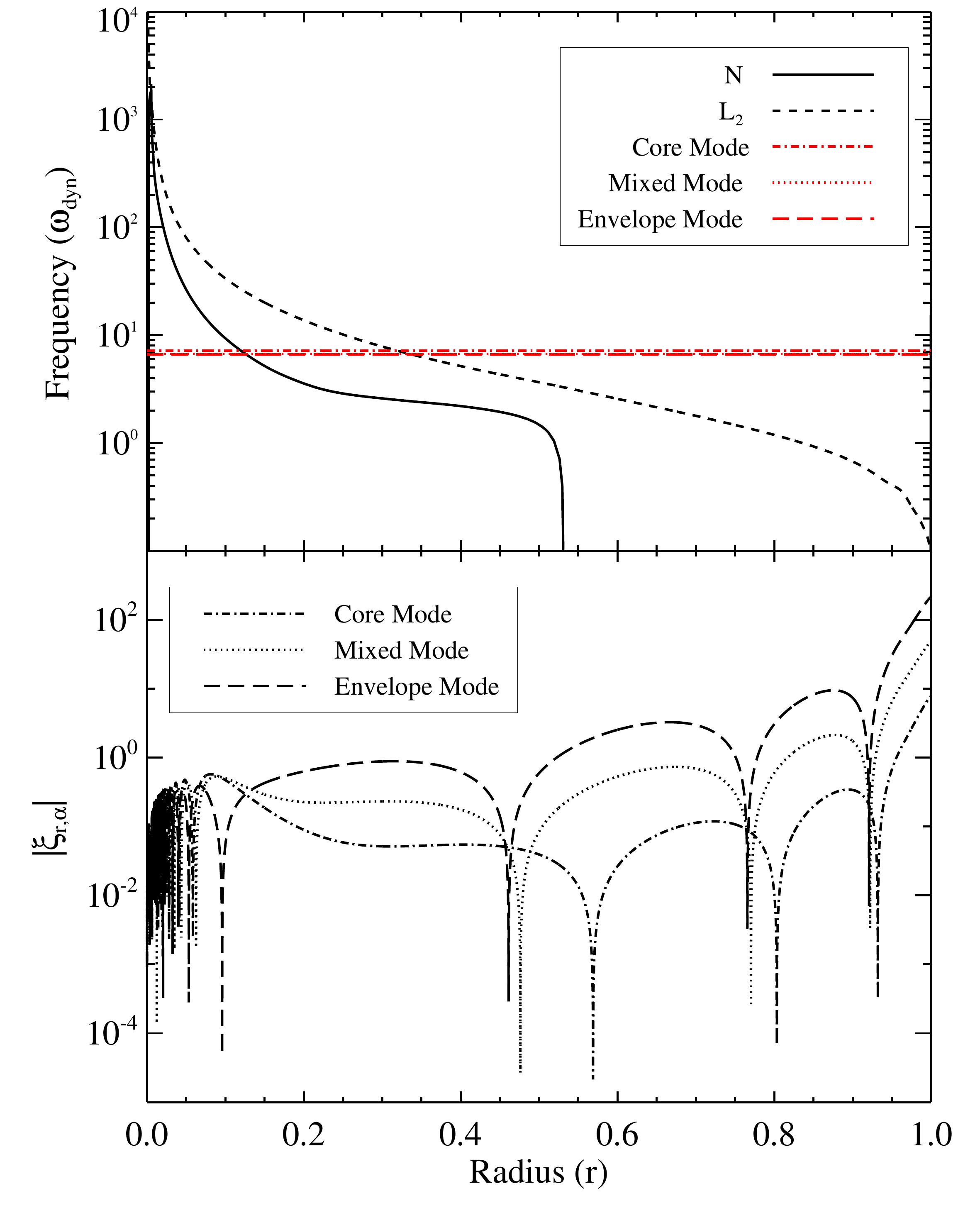}
\end{center}
\caption{ \label{struc} Top: the Brunt Vaisala frequency $N$ (solid line) and Lamb frequency $L_2$ (dashed line) as function of radius $r$. The red horizontal (overlapping) lines mark the eigenfrquencies $\bar{\omega}_\alpha$ of the modes shown in the bottom panel. Bottom: mode eigenfunctions $\xi_{r,\alpha}$ for an envelope mode with $\bar{\omega}_\alpha = 6.66$ (long dashed line), a neighboring mixed mode with $\bar{\omega}_\alpha = 6.71$ (dotted line), and a core mode with $\bar{\omega}_\alpha = 7.17$. All quantities are calculated for our $M= 3.0 M_\odot$, $R=12.4 R_\odot$ helium burning red giant model, and are plotted in units of $G=M=R=1$. The modes are normalized via equation (\ref{norm}).}
\end{figure*}

We generate a $M = 3.0 M_\odot$, $R = 12.4 R_\odot$, $T_{\rm eff} = 5100$K, $z=0.015$ helium burning red giant stellar model using the MESA stellar evolution code (Paxton et al. 2011). Figure \ref{struc} shows a propagation diagram for the stellar model. The high Brunt-Vaisala and Lamb frequencies in the radiative interior of the star allow high-order g-modes to propagate in the stellar interior, while the small Brunt-Vaisala and Lamb frequencies in the convective envelope of the star allow for p-mode propagation.

\begin{figure*}
\begin{center}
\includegraphics[scale=0.5]{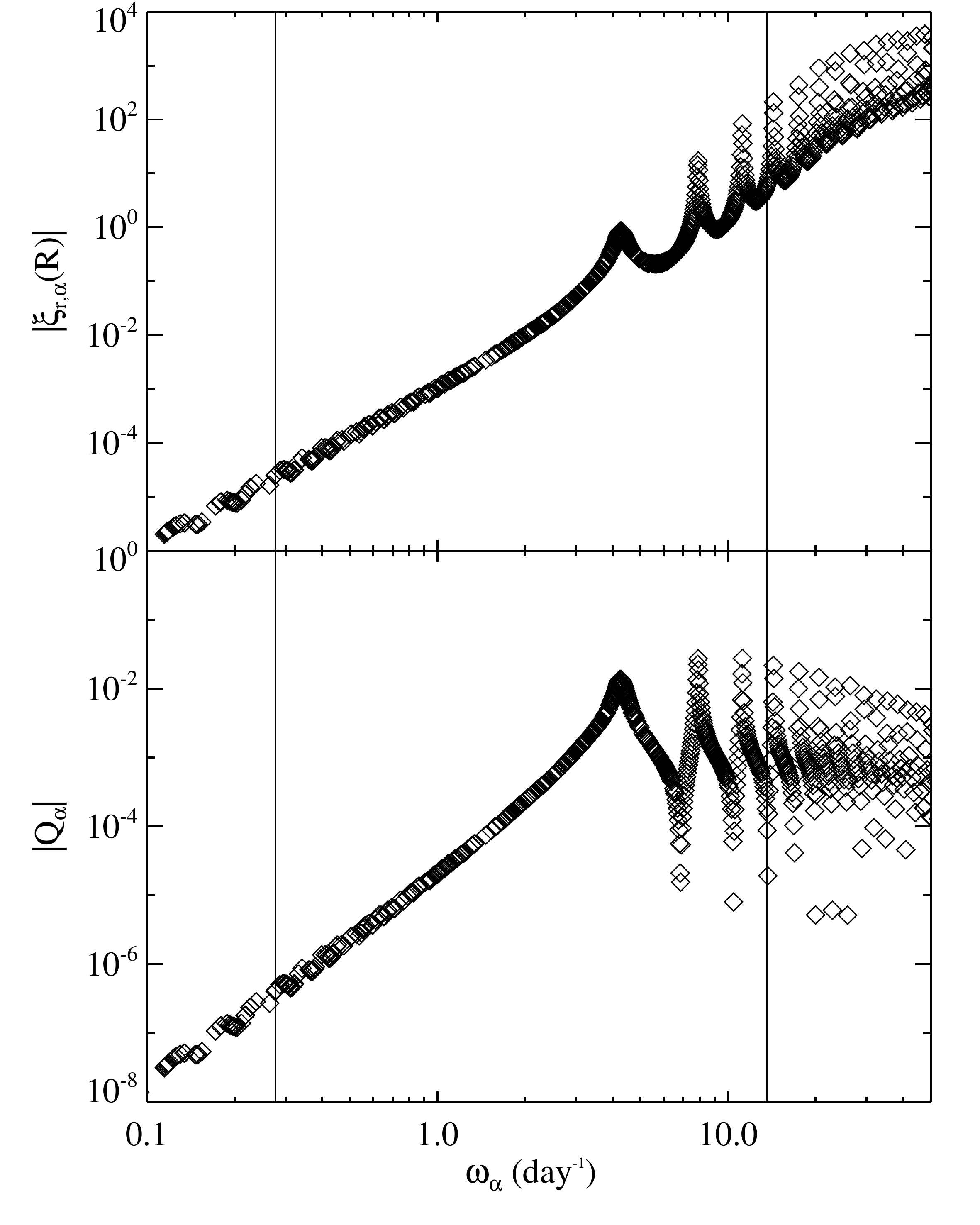}
\end{center}
\caption{ \label{Qxi} Radial displacement at the stellar surface $\xi_{r,\alpha}(R)$ (top) and overlap integral $Q_\alpha$ (bottom) for stellar oscillation modes as a function of mode eigenfrequency $\omega_\alpha$. The modes are calculated from our $M= 3.0 M_\odot$, $R=12.4 R_\odot$ red giant model. The modes at the tips of the spikes are envelope dominated mixed modes, while the surrounding modes are mixed modes whose inertia is split between the core and envelope. The vertical lines mark the values of $2\Omega_1$ and $2\Omega_{23}$ in the HD 181068 system.}
\end{figure*}

Using the model described above, we calculate the adiabatic stellar oscillation modes using the usual boundary conditions (see e.g., Unno et al. 1989), but normalized according to equation (\ref{norm}). Figure \ref{Qxi} shows the values of $\xi_{r,\alpha}(R)$ and $Q_\alpha$ as a function of $\omega_\alpha$ for our stellar model. Low frequency modes g-modes (lower than $\omega_\alpha \approx 1$) are trapped in the core. Higher frequency modes still have g-mode character in the core, but have p-mode character in the convective envelope of the star. Modes whose inertia lies primarily in the convective envelope are the envelope dominated mixed modes, while the inertia of the neighboring mixed modes is split between the core and envelope. Figures \ref{struc} and \ref{Qxi} demonstrate that the envelope modes have large values of $\xi_{r,\alpha}(R)$ and $Q_\alpha$ compared to neighboring mixed modes. Consequently, the envelope modes are easily excited to large amplitudes (because of the larger values of $Q_\alpha$) and produce large luminosity fluctuations [because of the larger values of $\xi_{r,\alpha}(R)$]. The envelope modes thus dominate the visible and energetic response of the star to high frequency tidal forcing.

\subsection{Comparison with Observed Luminosity Fluctuations}

Here, we compare our theory to observations of HD 181068. Unfortunately, it is difficult to predict the amplitude of a tidally excited mode, even if the system parameters are relatively well constrained. Part of the reason is that the value of the frequency detuning $D_{\alpha,m}$ (see equation \ref{D}) can vary by orders of magnitude over small changes in $\omega_\alpha$. Since the values of $\omega_\alpha$ depend on the precise mass, radius, and internal structure of the star, accurately calculating $D_{\alpha,m}$ for each mode is very difficult.

Instead, we choose to calculate the amplitude of the luminosity fluctuations as a function of the stellar radius $R_1$, which is constrained to be $R_1 = 12.46 \pm 0.15 R_\odot$. We compute the dimensionless values of $\bar{\omega}_\alpha$ for the stellar model described in Section \ref{model}, and then scale them to dimensional frequencies $\omega_\alpha$ using equation (\ref{baromega}). The range of radii $R_1$ is meant to encompass uncertainties in $\bar{\omega}_\alpha$, $M_1$, $R_1$, $\Omega_{s,1}$, non-adiabatic effects, etc., that affect the precise values of $\omega_\alpha$ in HD 181068.

\begin{figure*}
\begin{center}
\includegraphics[scale=0.55]{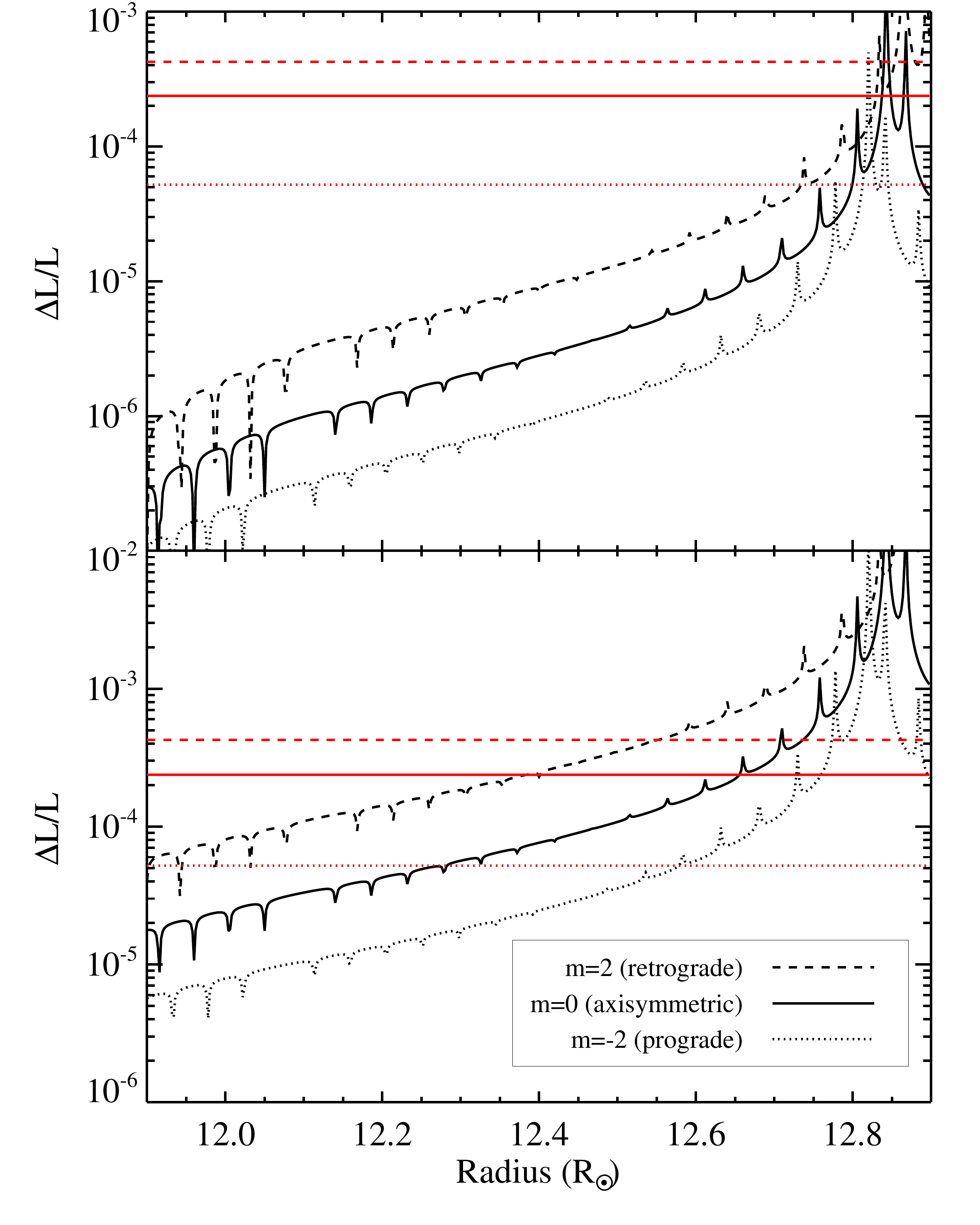}
\end{center}
\caption{ \label{vis} Luminosity variations $\Delta L/L$ as a function of stellar radius, using $H_\alpha=1$ (top panel) and $H_\alpha=\omega_\alpha$ (bottom panel), for our stellar model. The actual stellar model has $R_1=12.4 R_\odot$, but to make this plot we consider identical models scaled to different radii. The black lines are the luminosity variations of the mode at frequencies $2(\Omega_{23}-2\Omega_1)$ (dashed line), $2(\Omega_{23}-\Omega_1)$ (solid line), and $2\Omega_{23}$ (dotted line). The horizontal red lines are the observed luminosity variations in HD 181068 at the same frequencies.}
\end{figure*}

Figure \ref{vis} displays the theoretical and observed luminosity variations produced at the tidal forcing frequencies $2(\Omega_{23}-2\Omega_1)$, $2(\Omega_{23}-\Omega_1)$, and $2\Omega_{23}$. Recall that the observed variation at $2\Omega_{23}$ may be contaminated by imperfect eclipse subtraction and/or spotting effects. The luminosity variations are calculated using equation (\ref{deltaL}), using $\Omega_{23}$ and $\Omega_1$ observed in HD 181068, and using $M_2=M_3=0.9 M_\odot$ and $\theta_o = 87.5^\circ$ (as measured by Borkovits et al. 2012). The top panel uses $H_\alpha = 1$, while the bottom panel uses $H_\alpha=[l(l+1)/\bar{\omega}^2 - 4 - \bar{\omega}_\alpha^2]$. The luminosity variations have peaks and dips at values of $R_1$ for which the frequency of a mode is nearly resonant with the forcing frequency. For $H_\alpha = 1$, the theoretical luminosity variations are well below the observed variations, except very near a resonance with an envelope p-mode (i.e., at $R_1 \simeq 12.8R_\odot$). Unless the HD 181068 is in a resonance locking state (see Section \ref{orb}), it is unlikely to observe it so close to resonance. 

In contrast, the theoretical luminosity variations are closer to the observed variations for $H_\alpha=[l(l+1)/\bar{\omega}^2 - 4 - \bar{\omega}_\alpha^2]$. The best agreement is obtained for $R_1 \simeq 12.6 R_\odot$, and matches the luminosity variation for each theoretical and observed frequency to within a factor of 2. We conclude that the luminosity variations in HD 181068 are likely dominated by temperature effects, characterized by large values of $H_\alpha$. The ordering of the amplitudes of the oscillations is naturally explained by our theory, i.e., $\Delta L/L (m=2) > \Delta L/L (m=0) > \Delta L/L (m=-2)$, because $F_2 > F_0 > F_{-2}$. However, the measured amplitude ratios are slightly different than what we expect away from exact resonances, and the cause of the discrepancy is unclear.

For non-adiabatic modes, our theory also predicts the phases of the observed oscillations. In particular, equation (\ref{deltaL}) shows that for non-resonant adiabatic modes (i.e., $V_\alpha$ is real and $\psi_\alpha \simeq 0$ or $\psi_\alpha \simeq \pi$), the phase difference between two oscillation frequencies is $\Delta \phi \simeq m \phi_1$. This implies that the prominent frequencies $f_1$ and $f_2$ (which have $m=2$ and $m=0$, respectively) should be in phase at $\phi_1 \simeq 0$ and $\phi_1 \simeq \pi$, i.e., they should be in phase during the eclipses and occultations of Star 1. Indeed, the measured oscillations are in phase at these times, as can be seen from the simulated light curve in Figure \ref{beating}. This suggests that the observed luminosity fluctuations are being produced by non-resonant modes, consistent with our findings above.

%In HD 181068, $\Delta L/L$ for each oscillation frequency is observed to have a minimum near $\phi_{23} \simeq \phi_1 \simeq 0$, i.e., when Star 2 eclipses Star 3 and Star 1. This observation is consistent with the luminosity variations being produced by non-resonant [i.e., $\psi_{\alpha,m} \simeq 0$ or $\psi_{\alpha,m} \simeq \pi$, see equations (\ref{dL0}) and (\ref{dL2})], nearly adiabatic modes. It is also possible, but seemingly unlikely, that the phase shift in $\Delta L/L$ due to non-adiabatic and/or resonant effects is near a multiple of $\pi$. 

\section{Orbital Evolution of HD 181068}

\label{orb}

We now wish to calculate the orbital evolution induced by the tidally excited modes in realistic systems such as HD 181068. The  quantity in the orbital evolution equations (\ref{dL1dtm0b})-(\ref{dL23dtm2b}) with the most uncertainty is the damping rate $\gamma_\alpha$. An accurate calculation of $\gamma_\alpha$ requires a fully non-adiabatic calculation of mode eigenfrequencies, and must also include the turbulent damping of modes in the convection zone, which is not well understood. We choose to estimate $\gamma_\alpha$ using a quasiadiabatic WKB damping rate (see Burkart et al. 2012). We check that the damping rate calculated in this manner is the same order of magnitude as the damping rate inferred from observations of mode lifetimes of solar-like oscillations in red giants (see Belkacem 2012) for modes with $\omega_\alpha \approx \nu_{\rm max}$.

To understand the effects of tidally excited modes on the orbital evolution of a realistic triple system, we calculate the orbital evolution of a system resembling HD 181068 using equations (\ref{dL1dtm0b})-(\ref{dL23dtm2b}). We consider coplanar, circular orbits, as observed in HD 181068. The orbital frequencies change as
\be
\dot{\Omega}_1 = - \frac{3\dot{L}_1}{\mu_1 a_1^2}
\ee
and
\be
\dot{\Omega}_{23} = - \frac{3\dot{L}_{23}}{\mu_{23} a_{23}^2},
\ee
and the orbital semi-major axes change as $\dot{a}/a = -2\dot{\Omega}/(3\Omega)$. We define the tidal dissipation time scale as 
\be
\label{ttide}
t_{\rm tide} = \frac{a_{23}}{\dot{a}_{23}} = \frac{\mu_{23} a_{23}^2 \Omega_{23}}{2 \dot{L}_{23}}.
\ee 

The orbital evolution timescales due to tidally excited modes can be comparable to the stellar evolution time scales of the stars in the system. Hence, it is important to evolve the properties of the stars in the system simultaneously with the orbital elements. In our evolutions, we compute the radius of Star 1 as a function of time using the MESA stellar evolution code (Paxton 2011). The changing value of $R_1$ affects not only the values of $\dot{L}$ in equations (\ref{dL1dtm0b})-(\ref{dL23dtm2b}), but it also affects the eigenfrequencies $\omega_\alpha$ because the stellar oscillation frequencies scale as $\omega_\alpha \propto \sqrt{GM_1/R_1^3}$. Accounting for the changing eigenfrequencies is important because it can lead to resonance locking (Witte \& Savonije 1999, Fuller \& Lai 2012), allowing for tidal evolution on stellar evolutionary timescales rather than the longer non-resonant tidal evolution time scales.

In our calculations, we do not calculate new eigenfreqencies $\bar{\omega}_\alpha$ and associated eigenfunctions at each time step. We find that in the red giant phase of stellar evolution, the variations in $Q_\alpha$ as a function of time are relatively small, and that the variations in $\omega_\alpha$ are dominated by variations in $\omega_{\rm dyn}$ due to the changing value of $R_1$. We use the values of $\bar{\omega}_\alpha$ and $Q_\alpha$ shown in Figure \ref{Qxi}, and calculate $\omega_\alpha$ from equation (\ref{baromega}).

In addition to the tidally excited modes discussed in this paper, our evolutionary calculations should account for \textquotedblleft two-body" tidal effects, i.e., the tidal effects due to the zeroth order component of the tidal potential in equation (\ref{U0}). These tidal forces have no dependence on the small parameter $\epsilon$, and hence they act on much shorter time scales. Furthermore, tidal forces between Stars 2 and 3 will act on even shorter time scales due to their close separation. Therefore, two-body tidal forces cause Stars 2 and 3 to have a circular orbit around each other, rotating synchronously with that orbit. They also cause the center of mass of Stars 2 and 3 to have a circular orbit around Star 1, with Star 1 rotating synchronously with that orbit. Thus, in our evolution, we enforce $\Omega_{s,1} = \Omega_1$ and $\Omega_{s,2} = \Omega_{s,3} = \Omega_{23}$ at all times. We account for the angular momentum redistribution associated with these processes, although the stellar spins contain only a small fraction of the total angular momentum of the system.

Because the moment of inertia of Star 1 is much less than that of the orbit of Stars 2 and 3 about Star 1, a small amount of orbital angular momentum deposited in Star 1 by three-body tidal effects can drastically change its spin frequency. This angular momentum will then be transferred back to the orbit of Stars 2 and 3 about Star 1 by two-body tidal effects until synchronism is restored. Therefore, after each time step in our evolution, we calculate $\Delta L_1$, $\Delta L_{23}$ and $\Delta L_*$ from equations (\ref{dL1dtm0b})-(\ref{dLstardt}). We then adjust the values of $\Omega_{s,1}$ and $\Omega_1$ such that $\Omega_{s,1}=\Omega_1$ and the total angular momentum is conserved. Since $L_{*,1} \ll L_1$ for realistic parameters for a hierarchical triple system, the coupled tidal evolutions ensure that $\dot{L}_{1} \simeq -\dot{L}_{23}$. 

We also include orbital evolution due to induced eccentricity, magnetic braking, and gravitational radiation, as described in Appendix A. However, we find that the timescales associated with these processes are generally longer than the lifetime of a system such as HD 181068, so we do not discuss these effects in detail below.

\subsection{Results of Orbital Evolution}

\begin{figure*}
\begin{center}
\includegraphics[scale=0.55]{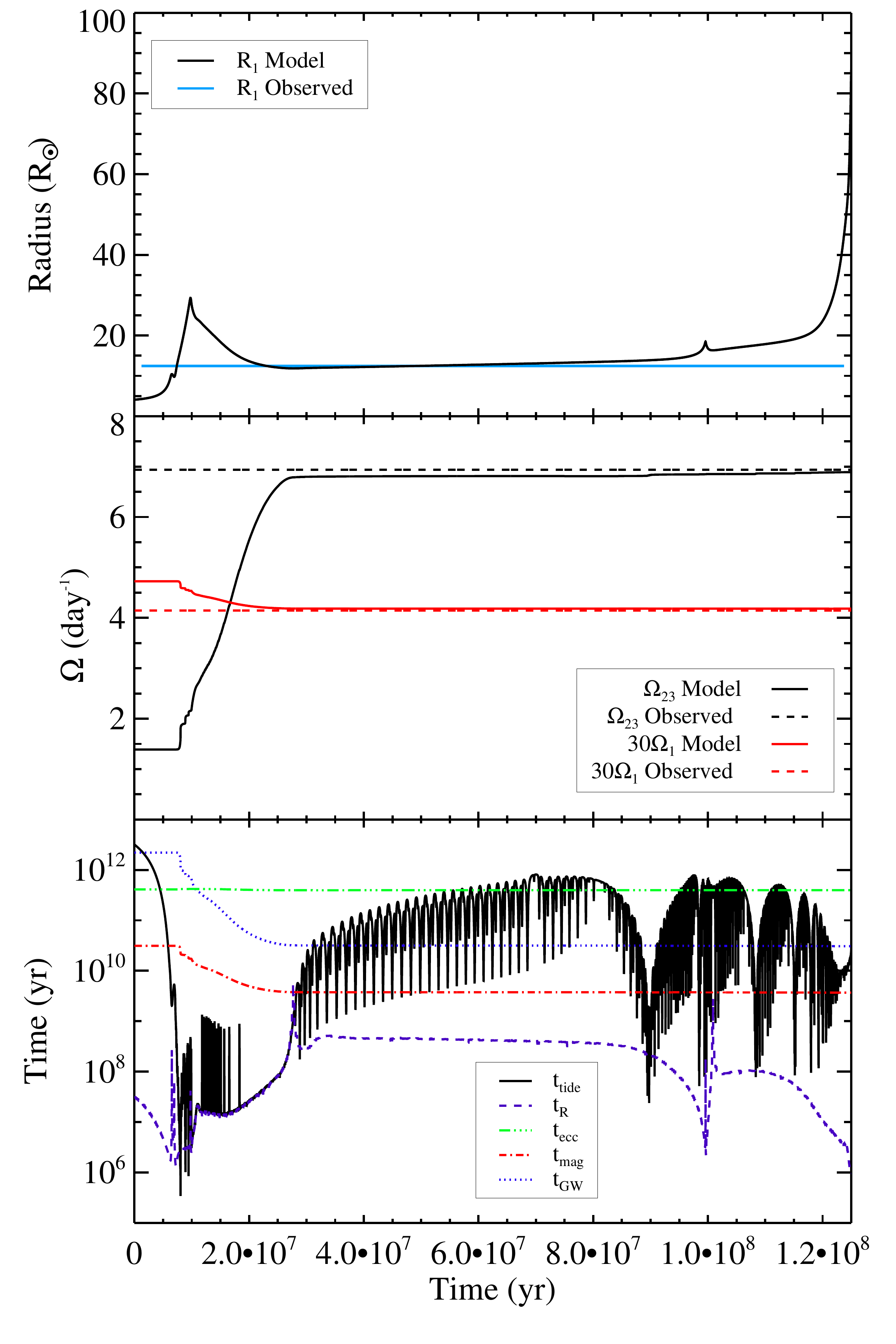}
\end{center}
\caption{ \label{evol} Top: Radius $R_1$ as a function of time since leaving the main sequence for an $M_1=3.0 M_\odot$ stellar model. The horizontal blue strip is the measured radius of HD 181068, within the measurement uncertainty. Middle: Angular orbital frequencies $\Omega_{23}$ (solid black line) and $10 \Omega_1$ (solid red line) as a function of time, including tidal dissipation. The horizontal lines are the values of $\Omega_{23}$ (dashed black) and $10 \Omega_1$ (dashed red) for the HD 181068 system. Bottom: tidal dissipation timescale $t_{\rm tide}$ (solid black line), induced eccentricity orbital decay time scale $t_{\rm ecc}$ (triple dot dashed green line), gravitational wave decay timescale $t_{\rm GW}$ (dotted blue line), magnetic braking orbital decay time scale $t_{\rm mag}$ (dot dashed red line), and stellar evolution time scale $t_R$ (dashed purple line) as a function of time.}
\end{figure*}

Figure \ref{evol} shows an example of our evolutionary calculations. We plot the stellar radius, $R_1$, orbital frequencies $\Omega_1$ and $\Omega_{23}$, and orbital decay time scales ($t_{\rm tide}$, $t_{\rm ecc}$, $t_{\rm mag}$, and $t_{\rm GW}$) as a function of time. We begin our orbital calculations as Star 1 is moving off the main sequence. The properties of the oscillation modes (calculated for helium burning red-giant model described in Section \ref{model}) are not appropriate for the initial model main sequence model, but become realistic as Star 1 moves up the red giant branch. We begin our calculation with orbital frequencies of $\Omega_{23,i} = 0.2 \Omega_{23,o}$ and $\Omega_{1,i} = 1.15 \Omega_{1,o}$, where $\Omega_{23,o}$ and $\Omega_{1,o}$ are the observed orbital frequencies in HD 181068. 

Let us start by examining the top panel of Figure \ref{evol}. The spike in radius at $t=1\times10^7$yr corresponds to the largest radius obtained during the red giant phase, while the long flat period between $2\times 10^7 {\rm yr} < t < 10^8 {\rm yr}$ is the core helium burning phase. Figure \ref{evol} indicates that Star 1 in HD 181068 could be ascending the red giant branch, but is most likely a horizontal branch star. The measured temperature of Star 1 is also consistent with these possibilities. 

The middle panel of Figure \ref{evol} shows that the orbits evolve substantially due to tidal dissipation. The outer orbital frequency, $\Omega_1$ becomes slightly smaller as angular momentum is transferred from the inner to the outer orbit. The inner orbital frequency $\Omega_{23}$ changes substantially, increasing by a factor of roughly five. The tidal dissipation begins as Star 1 moves up the red giant branch, increasing the value of $R_1$ and hence $\dot{L}_{23}$. Tidal dissipation also becomes much stronger during this stage because the frequencies of envelope p-modes (which couple strongly with the tidal potential) become comparable to the tidal forcing frequencies. The tidal dissipation remains strong as Star 1 shrinks and descends the red giant branch because the system enters into a resonance locking configuration (see Section \ref{reslock}). Once Star 1 settles onto the horizontal branch, the resonance locking ends and tidal dissipation is drastically reduced. 

The bottom panel of Figure \ref{evol} displays the relevant time scales of the evolution, including the stellar evolution time scale defined as
\be
\label{tr}
t_R = \frac{R_1}{|\dot{R}_1|}.
\ee
When Star 1 is on the main sequence, all the orbital time scales are longer than a Hubble time and can be ignored. As Star 1 moves up the red giant branch, the tidal dissipation time scale $t_{\rm tide}$ decreases by several orders of magnitude, as discussed above. When Star 1 is descending the red giant branch (at $t\approx 1.5 \times 10^7$yr), we see that $t_{\rm tide} \simeq t_R$. This is a natural consequence of resonance locking, which we discuss in Section \ref{reslock}.

In contrast, the value of $t_{\rm tide}$ is generally much larger during the core helium burning phase, due to the decreased values of $R_1$ and $a_{23}$. While in the core helium burning phase, the value of $t_{\rm tide}$ has sharp dips due to resonance crossings with mixed modes. There are also deeper, broader dips due to resonance crossings with envelope p-modes, such as the resonance crossing that occurs at $t \simeq 9 \times 10^7 {\rm yr}$. However, these resonance crossing events are fairly brief and produce only small amounts of tidal dissipation. Thus, we are thus unlikely to observe any orbital decay in HD~181068, unless the system is in a resonance-locking state.

The initial orbital frequencies were chosen such that the final values of $\Omega_{23}$ and $\Omega_1$ would be roughly equal to the observed orbital frequencies in HD 181068. This does not entail that our chosen values of $\Omega_{23,i}$ and $\Omega_{1,i}$ were the actual orbital frequencies of HD 181068 while Star 1 was on the main sequence. In reality, the initial orbital configuration of HD 181068 may have contained inclined or eccentric orbits for which the tidal dissipation rates may be substantially different. Nonetheless, the fact that a small value of $\Omega_{23,i}$ is required to match observations of HD 181068 indicates that the system may have experienced substantial orbital decay due to three-body tidal dissipation. 

In the future, HD~181068 will likely enter a common envelope phase as Star 1 evolves up the asymptotic giant branch and envelops Stars 2 and 3. Additionally, it is possible that HD~181068 reached its compact configuration through a mass transfer phase while Star 1 was on the red giant branch. Our MESA-generated stellar models indicate that our $3 M_\odot$ stellar model would not have overflown its Roche lobe (assuming $a_1$ was equal to its current value) while Star 1 was on the red giant branch. However, a $4 M_\odot$ star probably would have overflowed its Roche lobe while on the red giant branch, and it may be possible that this occurred in HD~181068. However, the outcome of stable mass transfer or common envelope evolution in triple systems is highly uncertain, and it warrants further study.

\subsection{Resonance Locking}
\label{reslock}

The orbital evolution discussed above contains periods of resonance locking in which a tidally excited mode is held near resonance for long periods of time, causing large amounts of tidal dissipation. Resonance locking involving tidally driven stellar oscillation modes was first investigated by Witte \& Savonije (1999) and recently proposed by Fuller \& Lai (2012) to explain the observed tidally driven oscillations in KOI-54. Resonance locking can occur when the frequency $\omega_\alpha$ of a mode changes due to a stellar evolutionary process. As the mode approaches resonance with a tidal forcing frequency, enhanced tidal dissipation occurs. The dissipation feeds back on the process, changing the value of the forcing frequency $\nu_f$ by spinning up the star or causing orbital decay. Under the right circumstances, the system maintains the nearly resonant configuration such that $\dot{\omega}_\alpha \simeq \dot{\nu}_f$, i.e., the orbital evolution timescale is roughly equal to the stellar evolution time scale. 

The resonance locking shown in Section \ref{orb} is qualitatively different from the locking investigated by Witte \& Savonije (1999) and Fuller \& Lai (2012), although the general principles described above are still true. In the case of a system like HD 181068, when Star 1 is descending the red giant branch, the frequencies of the stellar oscillation modes are increasing because the dynamical frequency of the star is increasing. When the value of $\omega_\alpha$ of an envelope mode is nearly resonant with a forcing frequency $\nu_f$, enhanced tidal dissipation occurs, causing the orbit of Stars 2 and 3 about one another to decay. Hence, the values of $\Omega_{23}$ and $\nu_f$ correspondingly increase, causing resonance locking.

When the system is resonantly locked, $\dot{\omega}_\alpha \simeq \dot{\nu}_f$. Assuming the star's oscillation frequencies change primarily due to the change in stellar radius, we have
\be
\label{doma}
\frac{\dot{\omega}_\alpha}{\omega_\alpha} \simeq - \frac{3}{2} \frac{\dot{R}_1}{R_1}.
\ee
Furthermore, the tidal forcing frequency changes primarily due to the increasing value of $\Omega_{23}$, so that
\be
\label{domf}
\frac{\dot{\nu}_f}{\nu_f} \simeq - \frac{3}{2} \frac{\dot{a}_{23}}{a_{23}}.
\ee
Therefore, during resonance locking, 
\be
\label{da}
\frac{\dot{R}_1}{R_1} \simeq \frac{\dot{a}_{23}}{a_{23}}, 
\ee
and thus $t_{\rm tide} \simeq t_R$. This explains the near equality of $t_{\rm tide}$ and $t_R$ during the descent of Star 1 from the red giant branch in Figure \ref{evol}. It also suggests that compact triples may endure a period of rapid tidal dissipation, caused by resonance locking, as the primary descends from the red giant branch toward the horizontal branch. Such tidal dissipation may lie in the near future for other observed compact hierarchical triples, such as KOI-126 (see Carter et al. 2011) and KOI-928 (see Steffen et al. 2011).

\section{Discussion}
\label{discussion}

We have demonstrated that a new tidal dissipation mechanism exists for stars in compact hierarchical triple systems, even after the system has reached the quasi-equilibrium state of aligned and circular orbits with aligned and synchronous stellar spins. The three-body tidal forces produce forcing at frequencies of $\sigma=2(\Omega_{23}-2\Omega_1)$, $\sigma=2(\Omega_{23}-\Omega_1)$, and $\sigma=2\Omega_{23}$ (although more forcing frequencies will exist for non-coplanar and non-circular orbits). If the primary star is a red giant with a large radius and thick convective envelope, the three-body tidal potential can couple strongly with the envelope p-modes of the star, exciting modes to large amplitudes.

We compare our results to \textit{Kepler} observations of HD 181068. The presence of oscillations in the lightcurve of HD 181068 at frequencies $\sigma=2(\Omega_{23} - 2\Omega_1)$, $\sigma=2(\Omega_{23} - \Omega_1)$, and $\sigma=2\Omega_{23}$ indicates that the primary exhibits stellar oscillation modes excited by the three-body tidal forcing described in this paper. The large amplitude of the oscillations in HD 181068 are either due to large temperature variations produced by small amplitude, non-resonant modes, or they are due to geometric distortions produced by large amplitude, nearly resonant modes. In the latter case, the tidal dissipation rate is rapid, and the modes may be locked in resonance (see Section \ref{reslock}). However, the amplitudes and phases of the modes are best explained by the non-resonant scenario. Furthermore, since the lifetime of the resonance locking phase is brief compared to the lifetime of the star on the horizontal branch, we find the non-resonant scenario to be much more likely. 

Because we have assumed adiabaticity when calculating stellar oscillation modes, there remains some uncertainty in the precise mode visibilities and damping rates. Fully non-adiabatic mode calculations can constrain the values of $H_\alpha$ and  $\gamma_\alpha$ in future studies. Furthermore, our calculations have been limited to the linear regime. As mentioned in Section \ref{forcing}, the three-body tidal forcing frequencies are nearly identical in the rotating frame of Star 1 if it is synchronized with the outer binary. This may allow for greatly enhanced non-linear mode coupling, affecting the mode amplitudes, damping rates, and visibilities.

The three-body tidal effects can also produce substantial orbital evolution among the stellar components, the main effect of which is the decay of the inner orbit of the compact binary. Our orbital evolution calculations for a system resembling HD 181068 reveal that, at most times, three-body tidal dissipation acts on long time scales and can be ignored. However, when the primary star is high on the red giant branch, three-body tidal dissipation may cause substantial orbital decay. Furthermore, stellar oscillation modes can become locked in resonance as the primary descends the red giant branch, resulting in greatly enhanced tidal dissipation. During resonance locking, tidal dissipation occurs on the same timescale as the stellar evolution, such that the orbital semi-major axis of Stars 2 and 3 decays as $\dot{a}_{23}/a_{23} \simeq \dot{R}_1/R_1$.

Future observations can detect and characterize tidally excited modes in compact triples. In non-eclipsing systems, the three-body nature of compact triples may not initially be detected from photometric and spectroscopic observations, especially if one star is much more luminous than its companions. The Kepler public red giant sample, consisting of more than 15,000 stars (Hekker et al. 2011), provides a unique resource that may contain several hidden clones of the Trinity system, which could be used to test our theory in a broader parameter space. The signature of three-body tidal forcing is a triplet of evenly spaced modes with frequencies $\sigma=2(\Omega_{23}-2\Omega_1)$, $\sigma=2(\Omega_{23}-\Omega_1)$, and $\sigma=2\Omega_{23}$, although the highest oscillation frequency may have a very small amplitude. The triplet could be mistaken for a rotationally split triplet of solar-like oscillations, but the tidally excited modes can be distinguished by their indefinitely long lifetimes (and thus narrow Fourier peaks). In low inclination systems, the $\sigma=2(\Omega_{23}-\Omega_1)$ oscillation will produce the largest luminosity fluctuations, while the $\sigma=2(\Omega_{23}-2\Omega_1)$ oscillation will be dominant in high-inclination systems like HD 181068.

\section*{Acknowledgments} 

We thank Dong Lai, Lars Bildsten, and Dan Tamayo for useful discussions. JF acknowledges the hospitality (Fall 2011) of the Kavli Institute for Theoretical Physics at UCSB (funded by the NSF through PHY 11-25915) where part of the work was carried out. This project has been supported by NSF grants AST-1008245 and AST-1211061, NASA grants NNX12AF85G and NNX10AP19G, the Hungarian OTKA Grants K76816, K83790 and MB08C 81013, and the ``Lend\"ulet-2009'' Young Researchers Program of the Hungarian Academy of Sciences. Part of the research leading to these results has received funding from the European Research Council under the European Community's Seventh Framework Programme (FP7/2007--2013)/ERC grant agreement n$^\circ$227224 (PROSPERITY). AD gratefully acknowledges financial support from the Magyary Zolt\'an Public Foundation. AD was supported by the Hungarian E\"otv\"os fellowship. AD has been supported by the J\'anos Bolyai Research Scholarship of the Hungarian Academy of Sciences. LLK wishes to thank support from the European Community’s Seventh Framework Programme (FP7/2007- 2013) under grant agreement number 269194. The Kepler Team and the Kepler Guest Observer Office are recognized for helping to make the mission and these data possible.

\def\apj{{Astrophys. J.}}
\def\apjs{{Astrophys. J. Supp.}}
\def\mnras{{Mon. Not. R. Astr. Soc.}}
\def\prl{{Phys. Rev. Lett.}}
\def\prd{{Phys. Rev. D}}
\def\apjl{{Astrophys. J. Let.}}
\def\pasp{{Publ. Astr. Soc. Pacific}}
\def\aapr{{Astr. Astr. Rev.}}

\appendix

\section{Non-tidal Orbital Evolution}

We wish to compare the effects of three-body tides to other physical mechanisms that may produce orbital evolution in triple-star systems. One such mechanism is tidal orbital decay due to the induced eccentricity of the inner binary (Stars 2 and 3) due to the perturbing gravitational influence of Star 1. This effect has been studied in detail by, e.g., Ford et al. 2000. In this mechanism, the eccentricity of the orbit of Stars 2 and 3 about one another oscillates around a small, non-zero value due to the presence of Star 1. Tidal interactions between Stars 2 and 3 act to re-circularize the orbit, dissipating energy and causing the orbit of Stars 2 and 3 about one another to decay. The orbital decay time scale due to dissipation in Star 2 is (Lithwick \& Wu 2012)
\be
\label{tecc}
t_{\rm ecc} = \frac{a_{23}}{\dot{a}_{23}} = \frac{1}{18\pi} \frac{Q}{k_2} \frac{\Omega_{23}R_2^3}{GM_2} \frac{1}{q(1+q)} \bigg(\frac{a_{23}}{R_2}\bigg)^8 \frac{1}{\langle e^2 \rangle},
\ee
where $Q$ is the tidal quality factor, $k_2$ is the constant of apsidal motion, $q=M_2/M_3$, and $\langle e^2 \rangle$ is the mean square eccentricity of the orbit of Stars 2 and 3 about one another. We have divided by an extra factor of two to account for tidal dissipation in both stars. According to Georgakarakos (2002), the mean square eccentricity for a nearly equal mass binary ($M_2 \simeq M_3$) is
\be
\label{ecc}
\langle e^2 \rangle \simeq \frac{43}{4} \bigg(\frac{M_1}{M_1\!+\!M_2\!+\!M_3}\bigg)^2 \bigg(\frac{\Omega_1}{\Omega_{23}}\bigg)^4.
\ee
However, for unequal mass inner binaries ($M_2 \neq M_3$) or systems near an orbital resonance, the mean square eccentricity may be significantly larger. In our calculations, we use $k_2=0.02$ (appropriate for low mass dwarf stars), and $Q=10^5$.

Another effect that can cause orbital decay in a compact triple system is magnetic braking. Magnetic braking is likely to act on short time scales in systems like HD 181068 because Stars 2 and 3 are rapidly rotating, low mass stars with convective envelopes that likely have strong magnetic dynamos. We estimate the torque due to magnetic braking according to the prescription of Krishnamurthi et al. 1997, yielding a magnetic braking timescale of
\be
\label{tmb}
t_{\rm mag} = \frac{a_{23}}{\dot{a}_{23}} = \frac{\mu_{23} a_{23}^2}{4 K_{MB} \Omega_{s,c}^2} \bigg(\frac{M_2}{M_\odot}\bigg)^{1/2} \bigg(\frac{R_2}{R_\odot}\bigg)^{-1/2},
\ee
where $\Omega_{s,c}$ is a critical rotation rate of $\Omega_c \sim 10 \Omega_{s,\odot}$, and $K_{MB}$ is a calibrated constant of $K_{MB} \approx 2.6 \times 10^{47} \rm{g \ s \ cm}^2$. Once again, we have divided by an extra factor of two to account for magnetic braking produced by Stars 2 and 3, assuming they are nearly identical. Star 1 may produce additional magnetic braking, but because $a_1 \gg a_{23}$ in hierarchical triples, the time scale of tidal orbital decay associated with this process is long, and can be ignored.

Finally, the orbit of Stars 2 and 3 may decay due to the emission of gravitational waves. The orbital decay time scale due to gravitational waves is
\be
\label{tgw}
t_{\rm GW} = \frac{a_{23}}{\dot{a}_{23}} = \frac{5c^5}{64G^3}\frac{a_{23}^4}{M_2 M_3 (M_2+M_3)}.
\ee


\begin{thebibliography}{}

\bibitem[]{}
Alexander, M.E., 1973, Astrophys. Space Sci., 23, 459

\bibitem[]{}
Baker, N. \& Kippenhahn, R., 1965, ApJ, 142, 868

\bibitem[]{}
Beck, P., et al., 2012, Nature, 481, 55

\bibitem[]{}
Bedding, T., et al., 2011, Nature, 471, 608

\bibitem[]{}
Borkovits, T., et al., 2012, MNRAS, accepted (arXiv:1210.1061)

\bibitem[]{}
Breger, M., et al., 1993, A\&A, 271, 482

\bibitem[]{}
Brown, T., Gilliland, R., Noyes, R., Ramsey, L., 1991, ApJ, 368, 599

\bibitem[]{}
Burkart, J., Quataert, E., Arras, P., Weinberg, N. 2011, MNRAS, 421, 983

\bibitem[]{}
Buta, R.J. \& Smith, M.A. 1979, ApJ, 232, 213

\bibitem[]{}
Carter, J., et al. 2011, Science, 331, 562

\bibitem[]{}
Chaplin, W., et al., 2011, Science, 332, 213

\bibitem[]{}
Chaplin, W., et al. 2011, ApJL, 732, 5

\bibitem[]{}
Correia, A., Laskar, J., Farago, F., Boué, G. 2011, CeMDA, 111, 105

\bibitem[]{}
Derekas, A., et al., 2011, Science, 332, 216

\bibitem[]{}
De Ridder, J., et al., 2009, Nature, 459, 398

\bibitem[]{}
Dziembowski, W. 1971, Acta Astronomica, 21, 3

\bibitem[]{}
Eggleton, P., Kisleva-Eggleton, L., 2001, ApJ, 562, 1012

\bibitem[]{}
Fabrycky, D., Tremaine, S., 2007, ApJ, 669, 1298

\bibitem[]{}
Ford, E., Kozinsky, B., Rasio, F., 2000, ApJ, 535, 385

\bibitem[]{}
Fuller, J., Lai, D. 2012, MNRAS, 420, 3126

\bibitem[]{}
Georgakarakos, N., 2002, MNRAS, 337, 559

\bibitem[]{}
Gouttebroze, P., Toutain, T., 1994, A\&A, 287, 535

\bibitem[]{}
Hekker, S., et al., 2011, MNRAS, 414, 2594

\bibitem[]{}
Houdek, G., Balmforth, N. J., Christensen-Dalsgaard, J., Gough, D. O., 1999, A\&A, 351, 582

\bibitem[]{} 
Huber, D., et al., 2010, ApJ,  723, 1607 

\bibitem[]{}
Huber, D., et al., 2011, ApJ, 743, 143

\bibitem[]{}
Hut, P. 1981, AA, 99, 126

\bibitem[]{} 
Kallinger, T., et al., 2010, A\&A, 522, A1 

\bibitem[]{}
Kumar, P. \& Quataert, E.J. 1998, ApJ 493, 412

\bibitem[]{}
Lai, D. 1996, ApJ, 466, L35

\bibitem[]{}
Lai, D. 1997, ApJ, 490, 847

\bibitem[]{}
Lenz, P., \& Breger, M., 2005, Comm. Asteroseis., 146, 53

\bibitem[]{}
Lithwick, Y. \& Wu, Y., 2012, ApJL, 756, 11

\bibitem[]{}
Mathur, S., et al., 2011, ApJ, 741, 119

\bibitem[]{}
Mazeh, T., Shaham, J., 1979, A\&A, 77, 145

\bibitem[]{}
Mosser, B., et al., 2012, arXiv:1209.3336

\bibitem[]{}
de Medeiros, J.R., Da Rocha, C., Mayor, M., 1996, A\&A 314, 499

\bibitem[]{}
Paxton, B., Bildsten, L., Dotter, A., Herwig, F., Lasaffre, P. Timmes, F. 2011, ApJS, 192, 3

\bibitem[]{}
Robinson, E., Kepler, S., Nather, R. 1982, ApJ, 259, 219

\bibitem[]{}
Samadi, et al., 2007, A\&A, 463, 297

\bibitem[]{}
Savitzky, A., Golay, M., 1964, Analytical Chemistry, 36, 1627

\bibitem[]{}
Schenck, A. K., Arras, P., Flanagan, E. E., Teukolsky, S. A., Wasserman, I. 2001, Phys Rev D, 65, 024001

\bibitem[]{}
Steffen, J., et al. 2011, MNRAS, 417L, 31

\bibitem[]{}
Thompson, S., et al. 2012, ApJ, 753, 86

\bibitem[]{}
Unno, W., Osaki, Y., Ando, H., Saio, H., Shibahashi, H. 1989, Nonradial Oscillations of Stars (University of Tokyo Press)

\bibitem[]{}
Welsh, W. F., et al. 2011, ApJS, 197, 4

\bibitem[]{}
Witte, M. G. \& Savonije, G.J. 1999, ApJ, 350, 129





\end{thebibliography}
\end{document}